\documentclass{elsart}
\usepackage[pdftex]{graphicx}
\usepackage{amsmath}
\usepackage{hyperref}
\usepackage{color}

\hypersetup{colorlinks,linkcolor=blue,citecolor=blue}

\newcommand{\figref}[1]{\figurename~\ref{fig:#1}}
\newcommand{\secref}[1]{Section~\ref{sec:#1}}
\renewcommand{\eqref}[1]{(\ref{eq:#1})}
\newcommand{\tabref}[1]{\tablename~\ref{tab:#1}}

\newcommand\de{\,{\mathrm d}} 

\newcommand{\PUC}{Y}

\newcommand{\PUCa}{a}
\newcommand{\PUCb}{b}
\newcommand{\PUCg}{g}
\newcommand{\PUCh}{h}

\newcommand{\PUCDx}{\Delta_1}
\newcommand{\PUCDy}{\Delta_2}
\newcommand{\PUCDz}{\Delta_3}

\newcommand{\tppF}{S}
\newcommand{\lpF}{L}

\newcommand{\bmath}[1]{\mbox{\boldmath$#1$}}
\newcommand{\vek}[1]{\bmath{#1}}
\newcommand{\mtrx}[1]{{\mathbf{#1}}}

\DeclareMathOperator*{\Argmin}{Argmin}

\newcommand{\imax}{i_{\rm max}}

\newcommand{\jmax}{j_{\rm max}}

\newcommand{\orig}[1]{\overline{#1}}
\newcommand{\weight}{\alpha}

\newcommand{\descr}{D} 
\newcommand{\phs}{p}   

\newcommand{\ww}{\sf{warp}}  
\newcommand{\wf}{\sf{weft}}  

\newcommand{\code}[1]{{\bf #1}}

\newcommand{\ME}{\vek{E}}
\newcommand{\mE}{\vek{\varepsilon}}

\newcommand{\MS}{\vek{\Sigma}}
\newcommand{\mS}{\vek{\sigma}}
\newcommand{\ML}{\mtrx{L}^{\sf H}}
\newcommand{\mL}{\mtrx{L}}
\newcommand{\trn}{{\sf ^T}}

\newcommand{\MH}{\vek{H}}
\newcommand{\mH}{\vek{h}}
\newcommand{\MQ}{\vek{Q}}
\newcommand{\mQ}{\vek{q}}

\newcommand{\avgs}[1]{\left\langle #1 \right\rangle} 

\newcommand{\volfrac}{\phi}

\newcommand{\newnot}[1]{#1}
\newcommand{\M}[1]{\newnot{\boldsymbol{#1}}}
\newcommand{\fl}{^{\newnot{*}}}

\journal{arxiv}

\begin{document}

\begin{frontmatter}

\title{Homogenization of plain weave composites with imperfect
microstructure: Part~II--Analysis of real-world materials}

\author[ctu]{Jan Vorel},
\author[ctu,it4i]{Jan Zeman},
\author[ctu]{Michal \v{S}ejnoha\corauthref{auth}},
\ead{sejnom@fsv.cvut.cz}
\corauth[auth]{Corresponding author. Tel.:~+420-2-2435-4494;
fax~+420-2-2431-0775}
\address[ctu]{Department of Mechanics, Faculty of Civil Engineering,
  Czech Technical University in Prague, Th\' akurova 7, 166 29 Prague
  6, Czech Republic}

\address[it4i]{Centre of Excellence IT4Innovations, V\v{S}B-TU Ostrava, 
17.listopadu 15/2172 708 33 Ostrava-Poruba, Czech Republic}

\begin{abstract}
A two-layer statistically equivalent periodic unit cell is offered to
predict a macroscopic response of plain weave multilayer carbon-carbon
textile composites. Falling-short in describing the most severe
geometrical imperfections of these material systems, the original
formulation presented in~\cite{Zeman:2004:HBPWI} is substantially
modified, now allowing for nesting and mutual shift of individual
layers of textile fabric in all three directions. Yet, the most
valuable asset of the present formulation is seen in the possibility
of reflecting the influence of meso-scale porosity through a system of
distorted voids.  Numerical predictions of both the effective thermal
conductivities and elastic stiffnesses provided through the
application of extended finite element method are compared with
available laboratory data and the results derived using the
Mori-Tanaka averaging scheme to support credibility of the present
approach, about as much as the reliability of local mechanical
properties found from nanoindentation tests performed directly on the
analyzed composite samples.
\end{abstract}


\end{frontmatter}

\section{Introduction}
Despite a significant progress in theoretical and computational
homogenization methods, material characterization techniques and
computational resources, the determination of overall response of
structural textile composites still remains an active research topic
in engineering materials science~\cite{Cox:2006:IQVT}. From a myriad
of modeling techniques developed in the last decades~(see e.g.  review
papers~\cite{Cox:1997:HAMTC,Chung:1999:WFC,Lomov:2007:MFET}), it is
generally accepted that detailed discretization techniques, and the
Finite Element Method~(FEM) in particular, remain the most powerful
and flexible tools available. The major weakness of these methods,
however, is the fact that their accuracy crucially depends on a
detailed specification of the complex microstructure of a
three-dimensional composite, usually based on two-dimensional
micrographs of material samples, e.g.~\cite[and reference
  therein]{Wentorf:1999:AMCW,Hivet:2005:C3D,Barbero:2006:FEM,Lomov:2007:MFET}.
Such a step is to a great extent complicated by \emph{random}
imperfections resulting from technological
operations~\cite{Pastore:1993:QPA,Yurgartis:1993:MYS}, which are
difficult to be incorporated to a computational model in a
well-defined way. If only the overall, or macroscopic response is the
important physical variable, it is sufficient to introduce structural
imperfections in a cumulative sense using available averaging schemes
such as the Voight and Reuss bounds~\cite{Yushanov:1998:STCM} or the
Mori-Tanaka method (MT)~\cite{Skocek:2007:MT}. When, on the other hand,
details of local stress and strain fields are required, it is
convenient to characterize the mesoscopic material heterogeneity by
introducing the concept of a Periodic Unit Cell~(PUC).

While application of PUCs in problems of strictly periodic media has a
rich history, their introduction in the field of random or imperfect
microstructures is still very much on the frontier, despite the fact
that the roots for incorporating basic features of random
microstructures into the formulation of a PUC were planted already in
mid 1990s in~\cite{Povirk:1995:IMI}. Additional extension presented
in~\cite{Zeman:2001:EPG}, see also our recent
work~\cite{Zeman:2007:FRM} for an overview, gave then rise to what we
now call the concept of Statistically Equivalent Periodic Unit
Cell~(SEPUC).  In contrast with traditional approaches, where
parameters of the unit cell model are directly measured from available
material samples, the SEPUC approach is based on their statistical
characterization. In particular, the procedure involves three basic
steps~\cite{Zeman:2007:FRM}:

\begin{itemize}

\item To capture the essential features of the heterogeneity pattern,
  the microstructure is characterized using appropriate statistical
  descriptors. Such data are essentially the only input needed for
  the determination of a unit cell.

\item A geometrical model of a unit cell is formulated and its key
  parameters are postulated. Definition of a suitable unit cell model
  is a modeling assumption made by a user, which sets the predictive
  capacities of SEPUC for an analyzed material system.

\item Parameters of the unit cell model are determined by matching the
  statistics of the complex microstructure and an idealized model,
  respectively. Due to multi-modal character of the objective
  function, soft-computing global optimization algorithms are usually
  employed to solve the associated problem.

\end{itemize}

It should be emphasized that the introduced concept is strictly based
on geometrical description of random media and as such it is closely
related to previous works on random media reconstruction, in
particular to the Yeong-Torquato algorithm presented
in~\cite{Yeong:1998:RRM,Yeong:1998:2D3D}. Such an approach is
fully generic, i.e. independent of a physical theory used to model the
material response. If needed, additional details related to the
simulation goals can be incorporated into the procedure without major
difficulties, e.g.~\cite{Bochenek:2004:RRM,Kumar:2006:UMR}, but of
course at the expense of computational complexity and the loss of its
generality.

In Part~I of this work~\cite{Zeman:2004:HBPWI}, the authors studied
the applicability of the SEPUC concept for the construction of a
single-layer unit cell reflecting selected imperfections typical of
textile composites. A detailed numerical studies, based on both
microstructural criteria and homogenized properties, revealed that
while a single-ply unit cell can take into account non-uniform layer
widths and tow undulation, it fails to characterize inter-layer shift
and nesting. In Part~II, we propose an extension of the original model
allowing us to address such imperfections, which have a strong
influence on the overall response of textile
composites~\cite{Woo:1997:EFTM,Jekabsons:2002:ESFL,Lomov:2003:NTL}.

\begin{figure}[ht]
\begin{center}
\includegraphics[width=12cm]{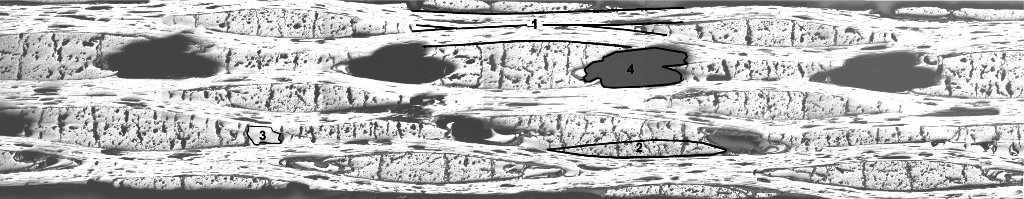}
\caption{Representative segment of eight-layer plain weave fabric laminate.}
\label{fig:meso_porosity}
\end{center}
\end{figure}

Such extensions, however, are hardly sufficient particularly in view
of a relatively high intrinsic porosity of C/C composites, which is in
the center of our current research efforts. It has been demonstrated
in our previous work~\cite{Tomkova:2008:IJMCE} that unless this
subject is properly addressed, inadequate results are obtained,
regardless of how ``exactly'' the geometrical details of the
meso-structure are represented by the computational
model. Unfortunately, the perceptible complexity of the porous phase
seen also in~\figref{meso_porosity} requires some
approximations. While densely packed transverse cracks affect the
homogenized properties of the fiber tow through a hierarchical
application of the Mori-Tanaka averaging scheme~\cite{Vorel:2008:SEM},
large inter-tow vacuoles (crimp voids), attributed to both
insufficient impregnation and thermal treatment, are introduced
directly into the originally void-free SEPUC in a discrete manner.
This step is addressed in~\secref{SEPUC_poros} together with the
finite element approximation based on X-FEM methodology as described
in~\cite{Sukumar:CMAME:2001,Moes:2003:CAHCMG}. It will be seen that
the resulting porous phase well approximates the true porosity
observed, e.g. via computational microtomography briefly mentioned
in~\secref{XrayEXP}.

Not only microstructural details but also properties of individual
composite constituents have a direct impact on the quality of
numerical predictions. Information supplied by manufacturers are,
however, often insufficient. Moreover, the carbon matrix of the
composite has properties dependent on particular manufacturing
parameters such as the magnitude and durations of the applied
temperature and pressure. Experimental derivation of some of the
parameters is therefore needed. In connection with the elastic
properties of the fiber and the matrix, the nanoindentation tests
performed directly on the composite are discussed in~\secref{NANOIND}
together with the determination of the necessary microstructural
parameters mentioned already in the previous paragraphs.

Still, most of the work presented in this paper is computational.  In
particular, a brief summary of the procedure for the determination of
the two-ply SEPUC for woven composites is given
in~\secref{SEPUC}. \secref{EXAMPLES} is then reserved for the
validation of the extracted geometrical and material parameters. To
that end, the heat conduction and classical elasticity homogenization
problems are compared with available experimental measurements. The
concluding remarks are presented in~\secref{CONCLUSION}.

\section{Experimental program}\label{sec:EXPERIMENT}
As already stated in the introductory part, much of the considered
here is primarily computational. However, no numerical predictions can
be certified if not supported by proper experimental
data~\cite{Knauss:IJSS:2000}. Considering the mesoscopic complexity of
C/C composites, the supportive role of experiments is assumed to have
the following four components:
\begin{itemize}
\item Two-dimensional image analysis providing binary bitmaps of the
  composite further exploited in the derivation of a two-layer SEPUC.
\item X-ray tomography yielding a three-dimensional map of
  distribution, shape and volume fraction of major pores to be
  introduced into a void-free SEPUC.
\item Nanoindentation tests supplying the local material parameters
  which either depend on the manufacturing process or are not
  disclosed by the producer.
\item Laboratory evaluation of effective properties both on micro and
  meso scales to support the experimental predictions. In the present
  study we adopt the experimental results available
  in~\cite{Cerny:Carbon:2000} for effective elastic moduli and
  in~\cite{Bohac:2005} for effective thermal conductivities,
  see~\secref{FEM}.
\end{itemize}

\begin{figure}[th]
\begin{tabular}{c@{\hspace{5mm}}c}
\includegraphics[width=.47\textwidth]{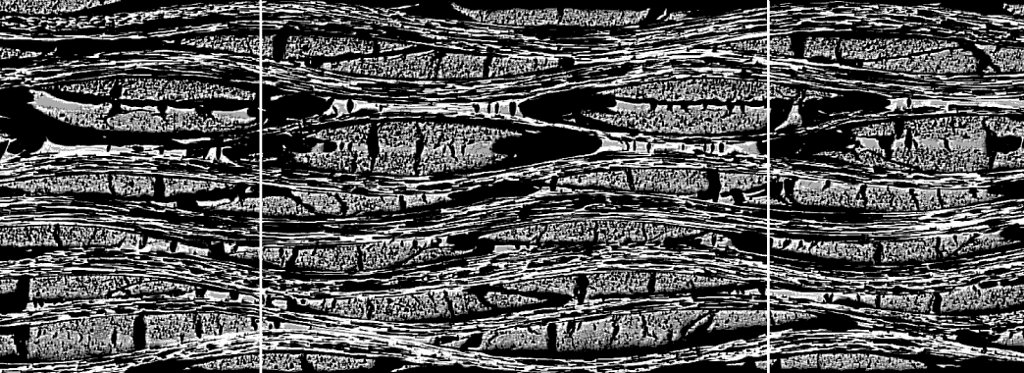} &
\includegraphics[height=0.171\textwidth,width=.475\textwidth]{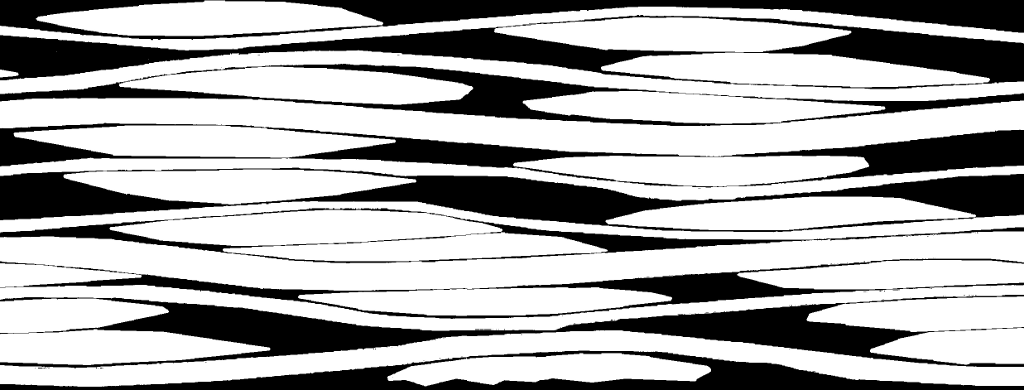} \\
(a)&(b) 
\end{tabular}
\caption{Examples of image analysis; (a)~multi-layered C/C composite, (b)~corresponding binary image}
\label{fig:img_anal_example}
\end{figure}

For the above purposes a carbon-polymer~(C/P) laminated plate was
first manufactured by molding together eight layers of carbon fabric
Hexcel G~1169 composed of carbon multifilament Torayca T 800 HB and
impregnated by phenolic resin Umaform LE. A set of twenty specimens
having dimensions $25 \times 2.5 \times 2.5$~mm were then cut out of
the laminate and subjected to further treatment (carbonization $C$ at
1000$^\circ$C, reimpregnation $I$, recarbonization, second
reimpregnation and final graphitization $G$ at 2200$^0$C $(CICICG)$)
to produce the carbon-carbon (C/C) composite. One particular
illustration together with a binary representative appears
in~\figref{img_anal_example}.

In the light of the first component of the experimental program, this
composite was thoroughly studied in~\cite{Tomkova:2004b,Tomkova:2006}
to yield basic structural parameters of a single-layer textile
composite summarized in~\tabref{parameters_SEPUC}.
\begin{table}[ht]
\caption{Parameters of the periodic unit cell~\cite{Tomkova:2004b}}
\centering
\begin{tabular}{lcccc}
\hline
\multicolumn{2}{l}{Parameter} &  Average & Standard deviation \\
\hline
Tow period &[$\mu$m] &             4500 & 300 \\
Tow height &[$\mu$m] &              150 &  20   \\
Inter-tow gap &[$\mu$m] &              400 & 105  \\
Layer height &[$\mu$m] &              300 &  50 \\
Horizontal shift &[$\mu$m] &   0 & 675 &  \\
Vertical shift &[$\mu$m]     &   0 & 110 \\
Porosity &[\%]      &   8 & 3.5 \\
Pore aspect ratio &[-]       & 0.4 & 0.2 \\
\hline
\end{tabular}
\label{tab:parameters_SEPUC}
\end{table}

\subsection{Three-dimensional X-ray microtomography}\label{sec:XrayEXP}
Porosity of C/C composites plays an important role in the derivation
of effective material properties~\cite{Tsukrov:MAMS:2005}. Although
sectioning is typically used to characterize porosity from a
two-dimensional images~\cite{Tomkova:2006}, the provided estimates of
the porosity may significantly pollute the final predictions of the
material response when compared to three-dimensional
simulations~\cite{Vorel:2008:SEM}. The X-ray microtomography, rendering
three-dimensional phase information directly, then becomes a valuable
tool.

Apart from usual medical diagnostic applications, radiologic imaging
is now commonly adopted in many other fields including material
research, archeology, biology and
other~\cite{Jirousek:NIMPR:11-2,Jirousek:JI:11-2}. For particular
applications to woven composites the reader is referred to
~\cite{Dobiasova:Carbon:2002,Djukic:2009:CE1,Djukic:2009:CE2}.

The experimental measurements were carried out at the Institute of the
Theoretical and Applied Mechanics, Academy of Sciences of the Czech
Republic, employing the Microfocus X-ray source L8601-01 (Hamamatsu
Photonics K.K.) with emission spot of $5~\mu$m and tungsten anode as
the source of X-rays.  For imaging, a large area X-ray detector
C7942CA-22 (Hamamatsu Photonics K.K.) with resolution $2368\times
2240$ pixels and physical dimensions $120\times 120$ mm was used. To
reduce the acquisition time $2\times 2$ pixel binning was adopted. The
scanning sequence consisted of $720$ scans with $0.5^{\circ}$ step.

\begin{figure}[ht]
\begin{center}
\begin{tabular}{c@{\hspace{5mm}}c}
\includegraphics*[width=0.48\textwidth]{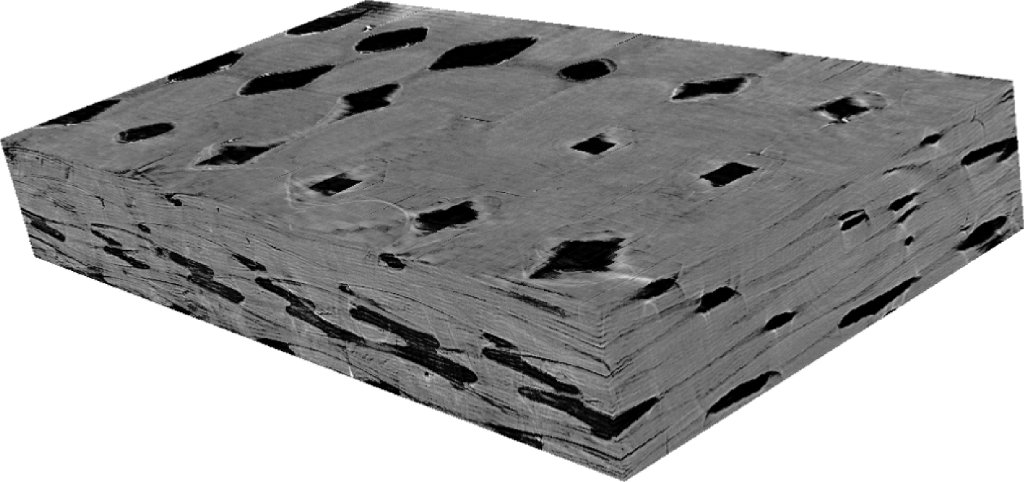}&
\includegraphics*[width=0.45\textwidth]{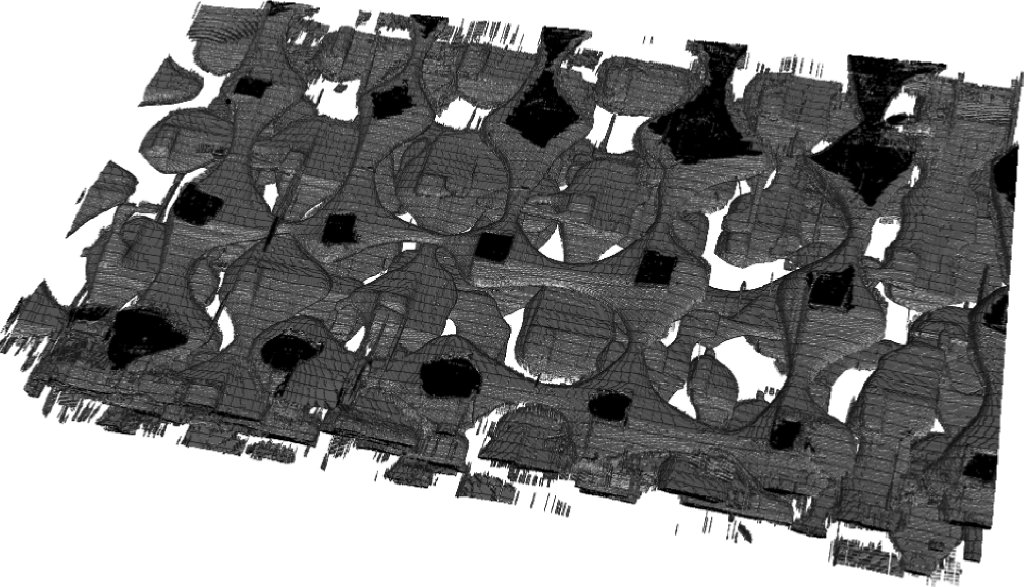}\\
(a)&(b)
\end{tabular}
\end{center}
\caption{(a) Reconstructed 3D
image of C/P composite, (b) reconstructed 3D image of porosity of C/C
composite}
\label{fig:xray-CCpore}
\end{figure}

A spatial resolution of the resulting 3D images was $12~\mu$m$^3$. A
particular example of the reconstructed 3D image of C/P multi-layered
plane-weave composite (green composite) together with the distribution
of major porosity derived for the carbonized ($CICIC$) sample is
presented in~\figref{xray-CCpore}.  Note that for the green
composite the major porosity amounted to $\approx 11\%$ while after
several steps of reimpregnation and carbonization it was reduced to
$\approx 8\%$, recall~\tabref{parameters_SEPUC}.

It is clear that these estimates are crucially dependent on the voxel
resolution with respect to the pore size. Nevertheless, if considering
the volume of an average pore equal to $100~\mu$m$^3$ and accept the
error of one voxel for each edge of the pore (a rectangular
parallelepiped is assumed for simplicity), the error induced by
calculation of the volume fraction of pores is less than $7\%$. Thus
for the carbonized sample the error ($0.07\times 8\%=0.56\%$) will not
exceed $1\%$ of the total volume.

\subsection{Phase elastic moduli from nanoindentation}\label{sec:NANOIND}

When predicting the effective behavior of composites from
computations, we generally rely on information about material
properties provided by the producer.  Unfortunately, these are often
insufficient for a three-dimensional analysis. It is also known that
material properties of the matrix much depend on the fabrication of
composite and may considerably deviate from those found experimentally
for large unconstrained material
samples~\cite{Dvorak:IJSS:2000}. Additional experiments, preferably
performed directly on the composite, are therefore often needed to
either validate the available local data or to derive the missing
ones.

At present, nanoindentation is the only experimental technique that
can be used for a direct measurement of mechanical properties at
material micro-level. A successful application of nanoindentation to
C/C composites has been reported
in~\cite{Kanari:Carbon:1997,Marx:Carbon:1999,Diss:Carbon:2002}. In the
present text, our attention is limited to the evaluation of the matrix
elastic modulus and the transverse elastic modulus of the fiber. The
remaining data are assumed to be estimated from those available in the
literature for similar material systems.

The reported measurements were performed at the Department of
Mechanics of the Czech Technical University in Prague adopting the CSM
Nanohardness tester. The testing head equipped with the Berkovich tip
was coupled with an optical microscope ($5\times$ and $100\times$
optical magnifications) and CCD camera. The loading range of this
equipment covers $0.1-500$ mN and the maximum penetration depth is
$20$ $\mu$m.

\begin{figure}[ht]
\begin{center}
\begin{tabular}{c@{\hspace{2mm}}c}
\includegraphics[angle=0,width=0.48\textwidth]{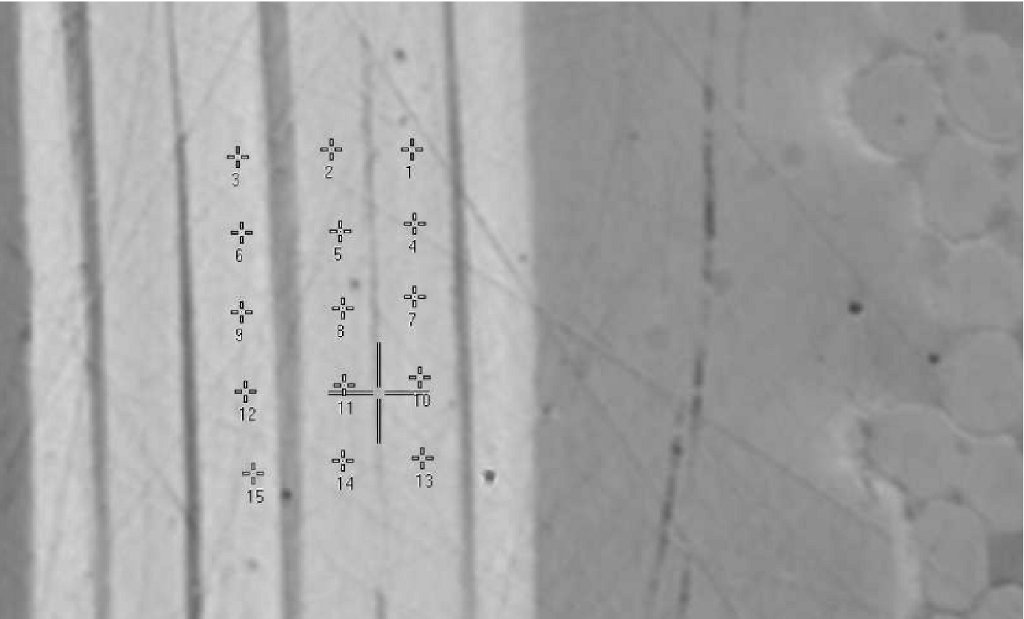}&
\includegraphics[angle=0,width=0.48\textwidth]{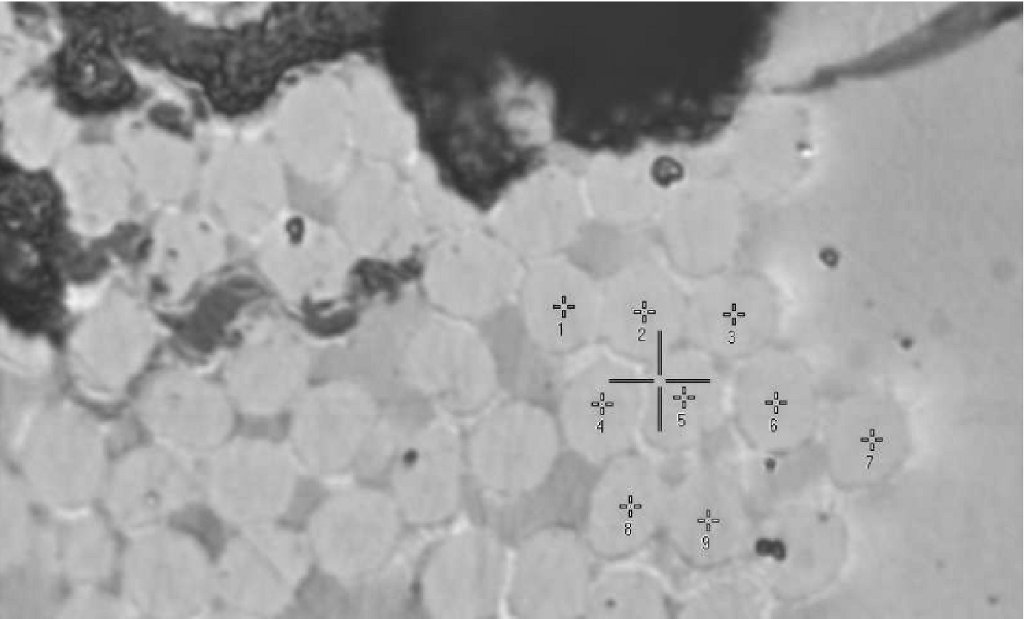}\\
(a)&(b)\\
\end{tabular}
\end{center}
\caption{Nanoindentation - location of indents: (a)~transverse direction,
(b)~perpendicular direction (compression)}
\label{fig:nano-sam}
\end{figure}

Three locations, distinctly separated in optic microscope, were tested
- matrix, parallel fibers \figref{nano-sam}(a), perpendicular fibers
\figref{nano-sam}(b). The matrix was therefore assumed isotropic and
possible anisotropy, which may arise inside the fiber tow, was not
considered. As seen in \figref{nano-sam} several measurements were
recorded for each of the three locations.

\begin{figure}
\begin{center}
\begin{tabular}{c@{\hspace{1mm}}c}
\includegraphics*[angle=0,width=0.48\textwidth]{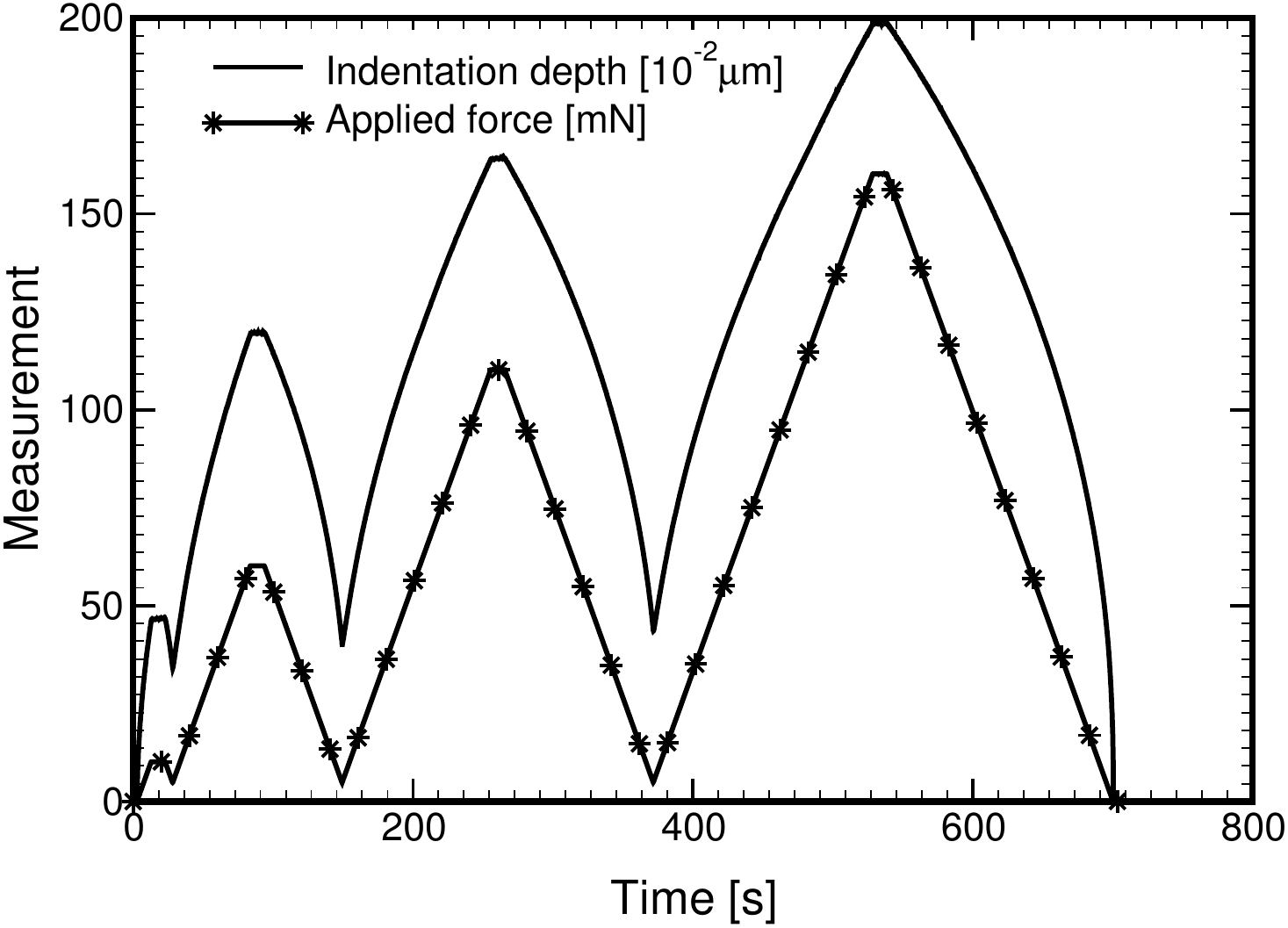}&
\includegraphics*[angle=0,width=0.48\textwidth]{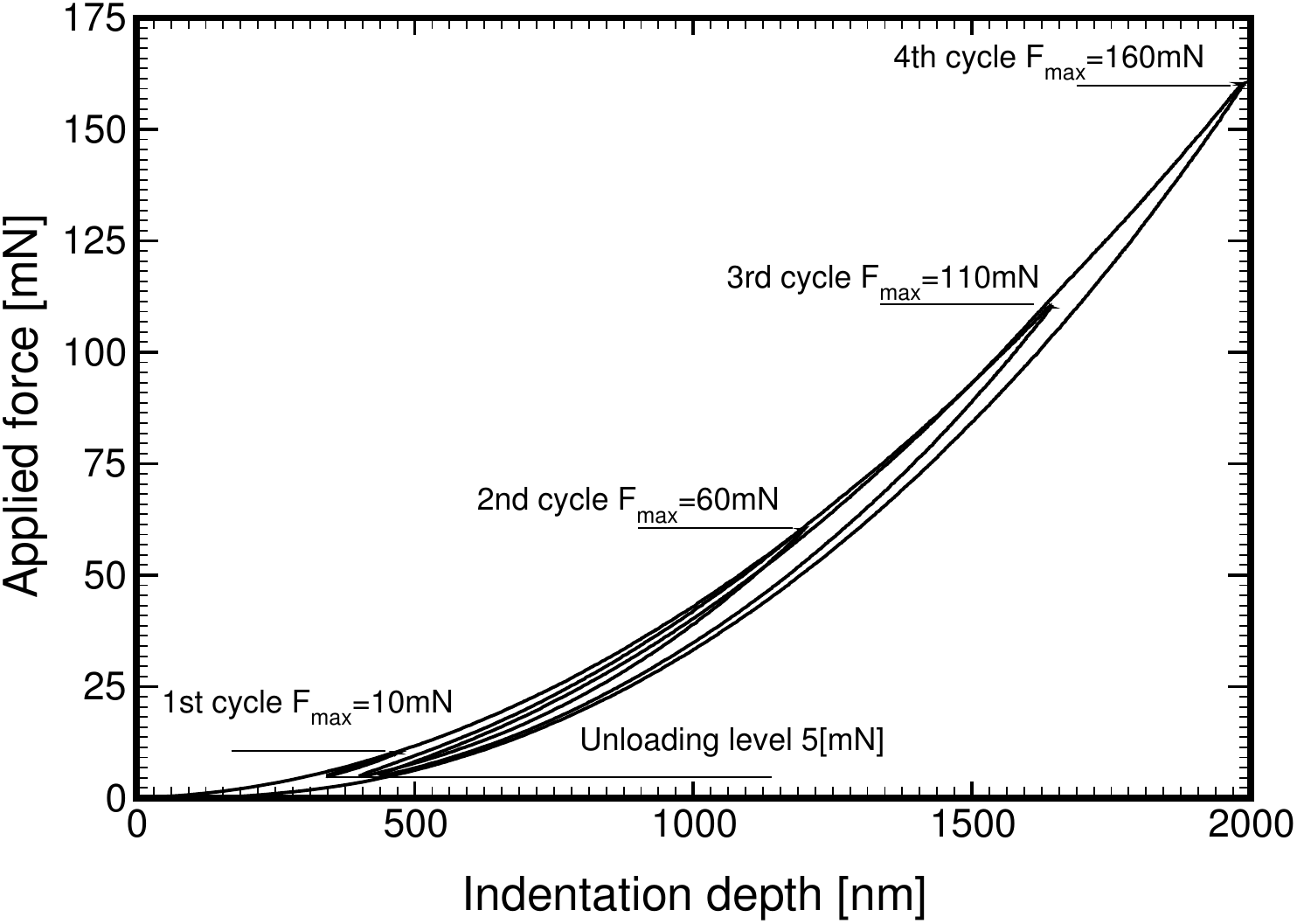}\\
(a)&(b)
\end{tabular}
\end{center}
\caption{Measurements of matrix properties: (a)~time variation of load and indentation depth, (b)~indentation curves}
\label{fig:nano-res-matrix}
\end{figure}

\begin{figure}[ht]
\begin{center}
\begin{tabular}{c@{\hspace{1mm}}c}
\includegraphics*[angle=0,width=0.48\textwidth]{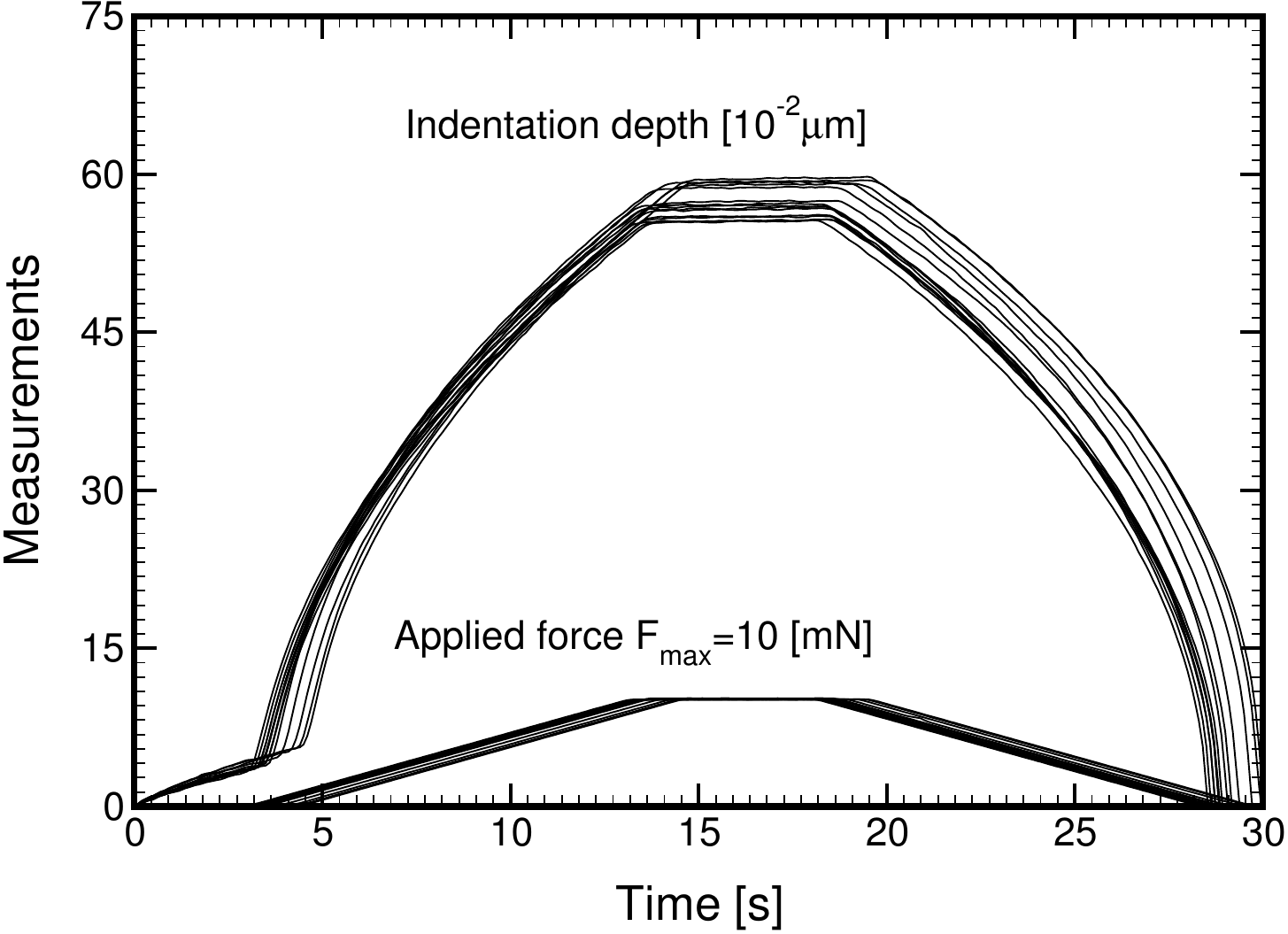}&
\includegraphics*[angle=0,width=0.48\textwidth]{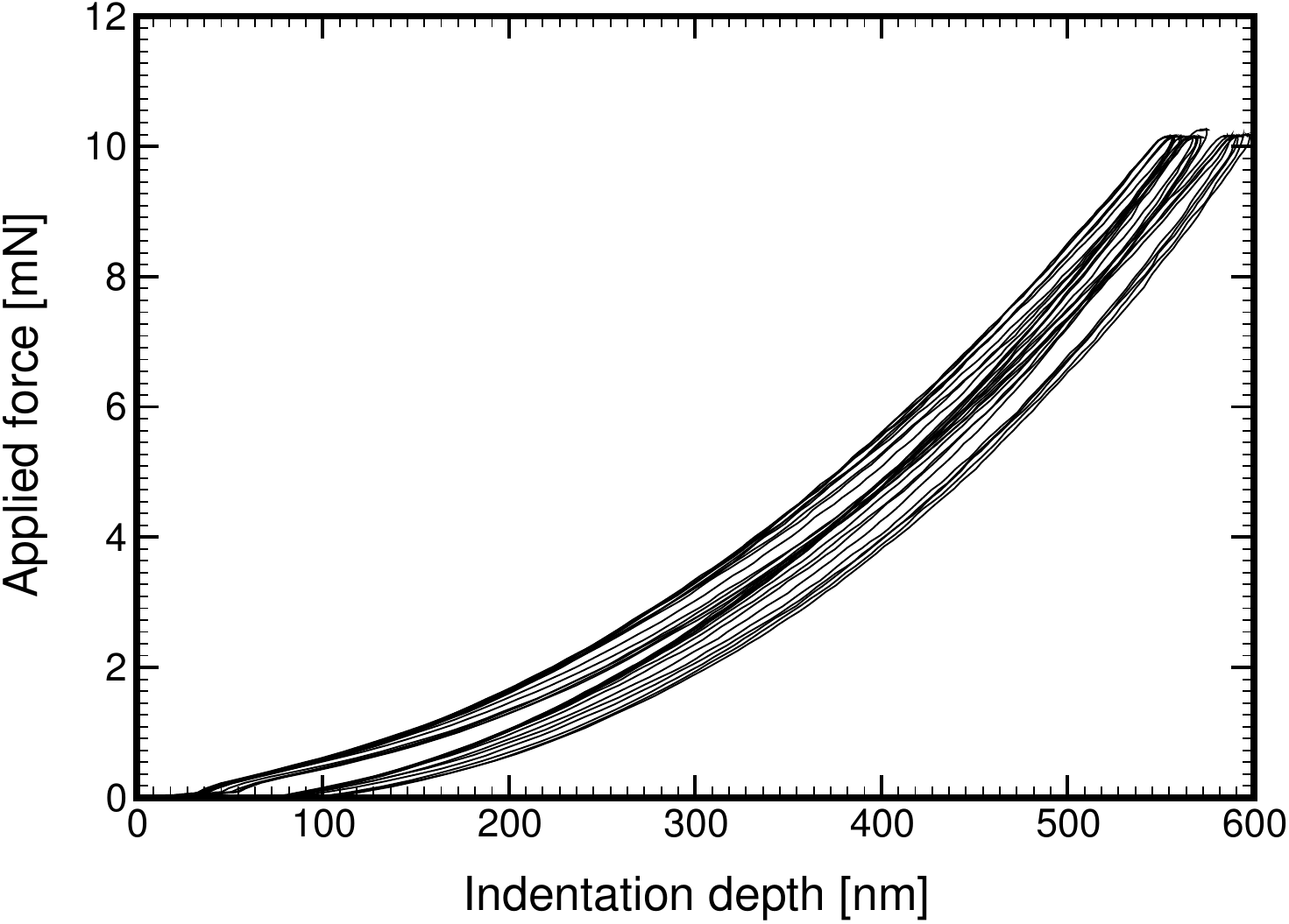}\\
(a)&(b)
\end{tabular}
\end{center}
\caption{Measurements of fiber properties: (a)~time variation of load and indentation depth, (b)~indentation curves}
\label{fig:nano-res-fiber}
\end{figure}

The loading sequence for the matrix phase appears
in~\figref{nano-res-matrix} showing the results for one particular
indent, whereas for the fiber phase it is plotted
in~\figref{nano-res-fiber} displaying several different indents.

As evident from Figs.~\ref{fig:nano-res-matrix}(a) and
~\ref{fig:nano-res-fiber}(a), no creep was observed for both the
matrix and and fibers as can be judged from no increase of an
indentation depth for the period of constant load. Also note that
the relative shift of individual indentation curves in
\figref{nano-res-fiber} caused by the difference of the onset of
actual measurements is irrelevant (the measuring device introduces a
non-zero indentation even before establishing a contact between the
indenter and the measured sample, see the initial branches of the time
variation of load and indentation depth in
\figref{nano-res-fiber}(a)). The initial slope of the unloading part
of indentation curves on the other hand plays a significant role when
assessing the quality of measurements, since this branch is adopted to
extract the elastic moduli~\cite{Nemecek:MCH:09} using the well known
Oliver-Pharr procedure~\cite{Oliver:JMR:1997}. In this case, the
unloading branch is almost identical for all measurements.

The complete set of material parameters, both measured averages
labeled by $^*$ and those adopted from the literature, is available
in~\tabref{param_CC}. Note that the matrix modulus agrees relatively
well with the one found for the glassy carbon
in~\cite{Diss:Carbon:2002}.

\begin{table}[ht]
\caption{Material parameters of individual phases}
\centering
\begin{tabular}{clccc}
\hline
&& Young modulus & Shear modulus & Poisson ratio   \\
\multicolumn{2}{c}{Material} & [GPa] & [GPa] & [-] \\
\hline 
fiber  & longitudinal       & 294  & 11.8 & 0.24 \\
       & transverse         & 14.7$^*$ & 4.1  & 0.4  \\
\hline
\multicolumn{2}{c}{matrix}  & 23.6$^*$ & 9.8 & 0.2  \\
\hline
\end{tabular}
\label{tab:param_CC}
\end{table}

\section{Statistically equivalent period unit cell}\label{sec:SEPUC}
The concept of Statistically Equivalent Periodic Unit Cell (SEPUC) for
random or imperfect microstructures is now well
established. Individual steps, enabling the substitution of real
microstructures by their simplified artificial representatives - the
SEPUCs - are described,
e.g. in~\cite{Zeman:2001:EPG,Zeman:2004:HBPWI,Zeman:2007:FRM} and
additional references given below.  Herein, these steps are briefly
reviewed concentrating on the specifics of multi-layer woven
composites.

\subsection{Geometrical mesostructural model}\label{sec:SEPUC_model}

\begin{figure}[th]
\centering
\begin{tabular}{c@{\hspace{5mm}}c}
\includegraphics[width=.42\textwidth]{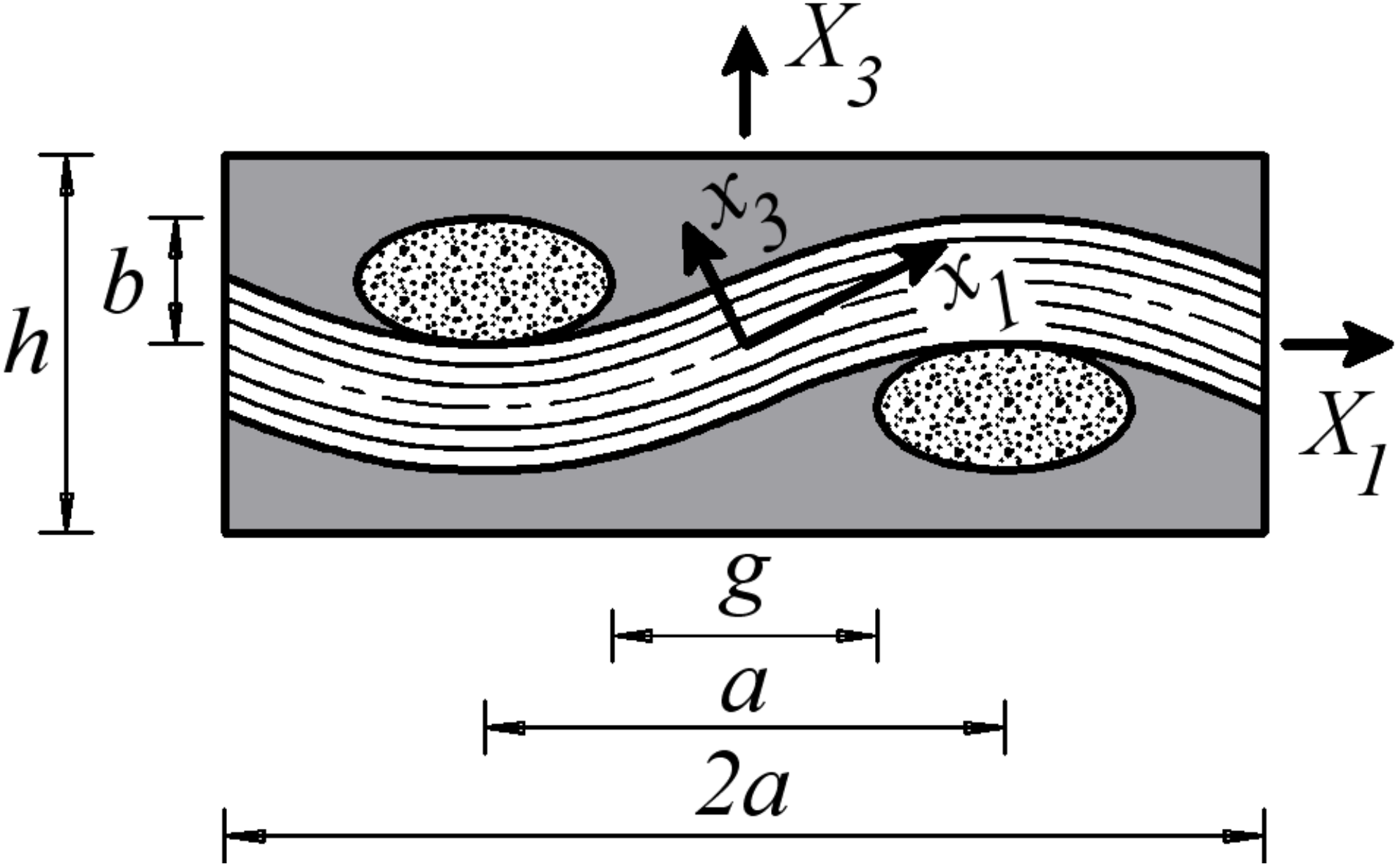} &
\includegraphics[width=.42\textwidth]{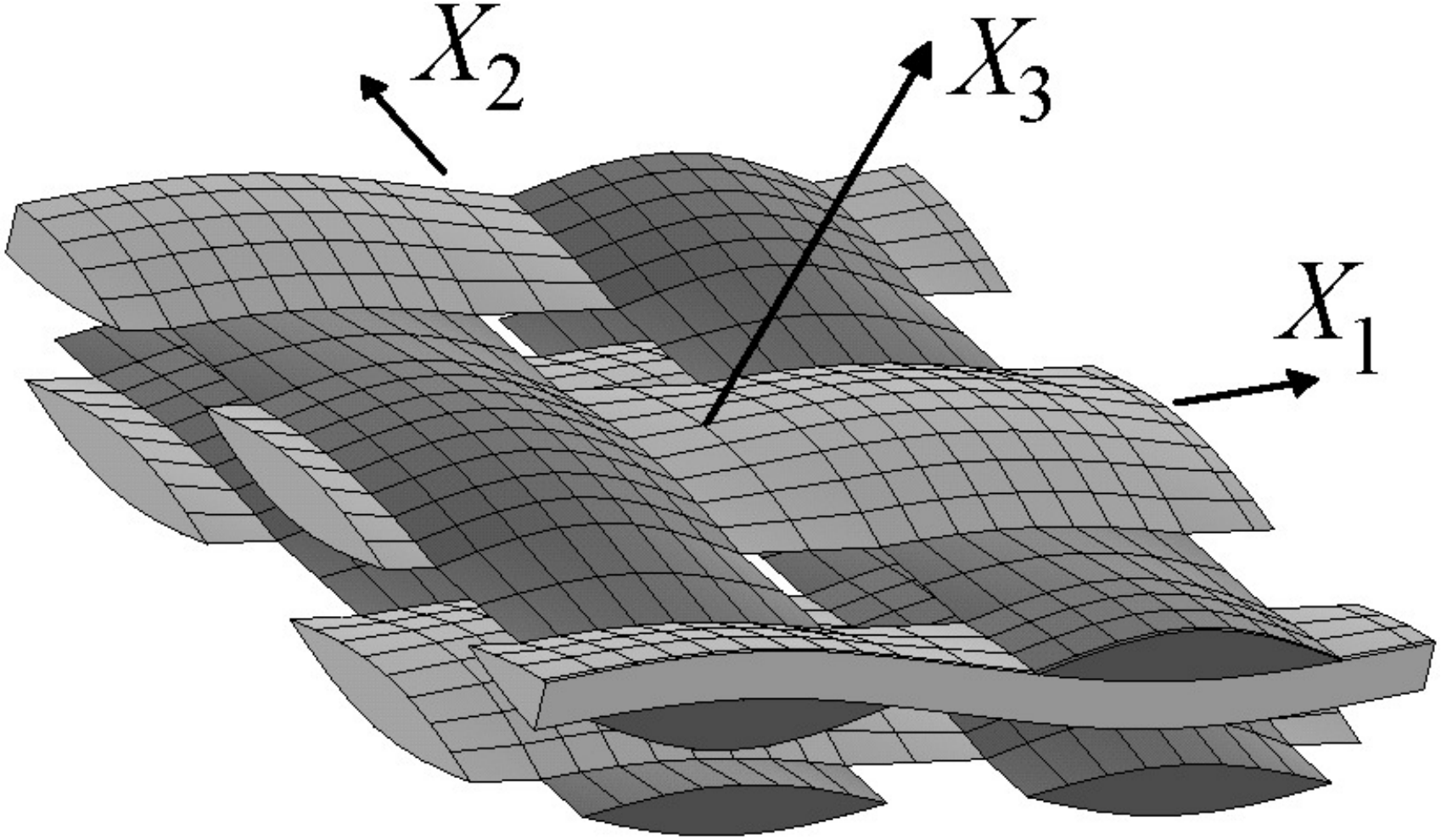} \\
(a)&(b)
\end{tabular}
\begin{tabular}{c}
\includegraphics[width=.55\textwidth]{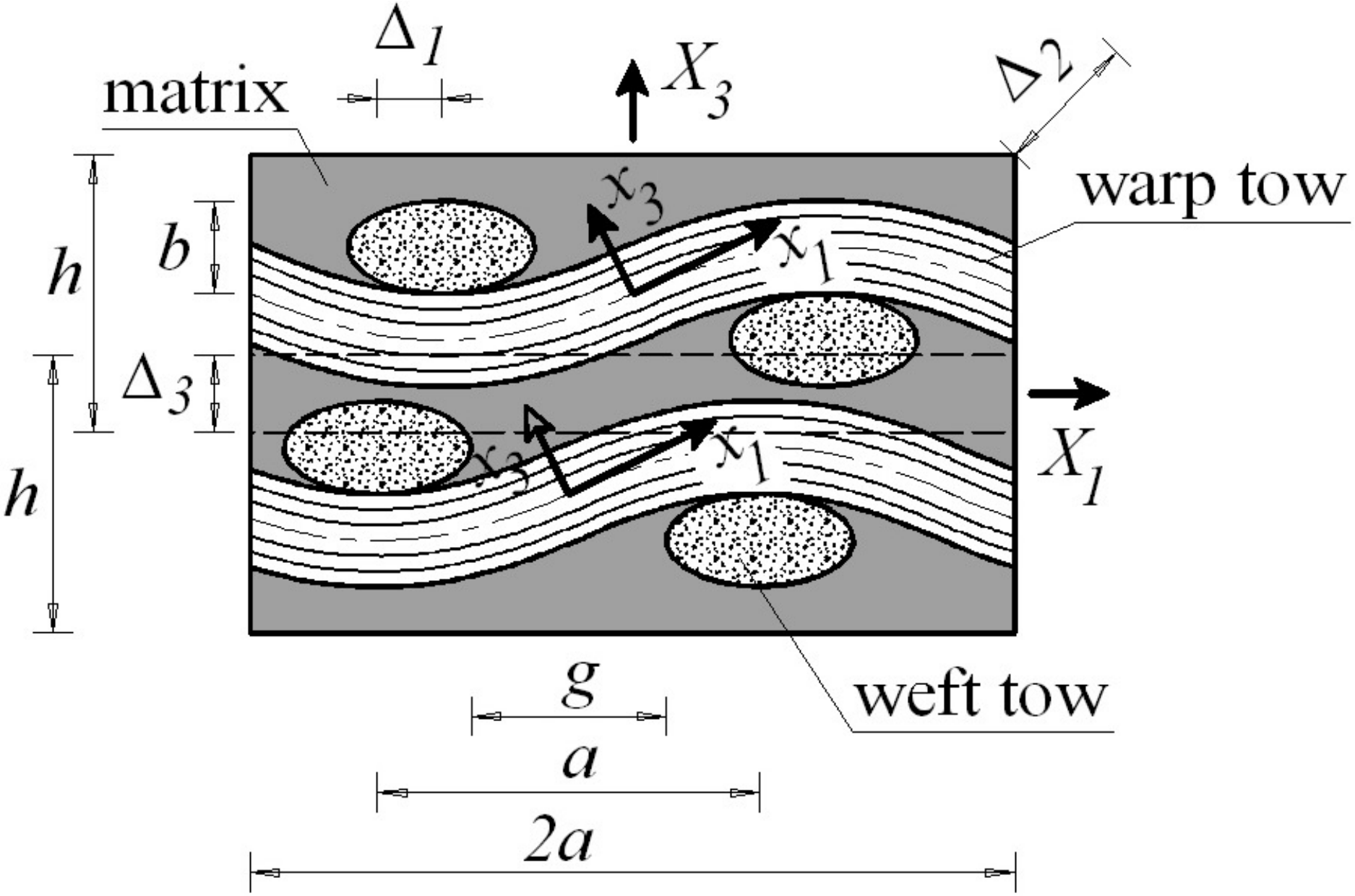}\\
(c)
\end{tabular}
\caption{Geometrical model of SEPUC; (a)~Two-dimensional cut of a one-layer model,
  (b)~two-layer model including periodic extension of upper layer,
  (c)~two-dimensional cut}
\label{fig:SEPUC}
\end{figure}

The basic building block of the adopted SEPUC is provided by a
single-ply model of plain weave composite geometry proposed by Kuhn
and Charalambides in~\cite{Kuhn:1999:MPWCG}. The model consists of two
orthogonal warp and weft tows embedded in the matrix phase and it is
parametrized by four basic quantities, directly measurable by
two-dimensional image analysis~(recall \tabref{parameters_SEPUC}): the
half-period of tow undulation~$\PUCa$, the maximal tow
thickness~$\PUCb$, the width of the intra-tow gap~$\PUCg$ and the
overall height of the ply~$\PUCh$, see~\figref{SEPUC}~(a). The
three-dimensional woven composite SEPUC, shown in~\figref{SEPUC}~(b),
is formed by two identical one-layer blocks, relatively shifted by
distances $\PUCDx$, $\PUCDy$ and $\PUCDz$ in the direction of the
corresponding coordinate axes. Finally, cutting a SEPUC by the plane
$X_2=\pm\PUCa$ or $X_1 = \pm\PUCa$ yields the warp or weft
two-dimensional sections, used as the basis for the determination of
the unit cell parameters.

\subsection{Quantification of random microstructure}\label{sec:SEPUC_quantif}
\begin{figure}[th]
\centering
\includegraphics{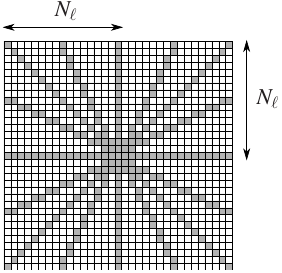}
\caption{Example of sample template for lineal path function generated
  by $N_d = 16$~segments, each of which is discretized using $N_\ell =
  17$~pixels.}
\label{fig:template}
\end{figure}

%
Assuming a statistically homogeneous and ergodic binary material
system, two basic statistical functions are available to capture
essential characteristics of the analyzed tow-matrix material
system. The first descriptor is a two-point probability function
$\tppF(\vek{X})$~\cite{Torquato:1982:MTP}, which quantifies the
probability of two points, separated by a vector $\vek{X}$, being both
found in the domain occupied by the warp and weft tows. The
alternative statistics, proposed by Lu and Torquato~\cite{Lu:1992:LPF}
to capture long-range effects, is the linear path function
$\lpF(\vek{X})$ giving the probability that a randomly placed segment
$\vek{X}$ is fully contained in the tow region. Both descriptors can
be easily computed for digitized microstructures; in particular, the
Fast Fourier transform library \code{FFTW}~\cite{Frigo:2005:FFTW3} is
used to evaluate the $\tppF$ function and the sampling template
consisting of $N_d$ concentric rays discretized by $N_\ell$
pixels~(cf.~\figref{template}) is employed to determine the linear
path function. The periodic boundary conditions were adopted
for both descriptors to eliminate edge
effects~\cite{Gajdosik:2006:QAFC}.

\subsection{Calibration of SEPUC parameters}\label{sec:SEPUC_calib}
%
In overall, the adopted model of the unit cell involves seven
independent parameters
\begin{equation}
\vek{y} = 
\left[ 
  \PUCa, \PUCb, \PUCg, \PUCh, 
  \PUCDx, \PUCDy, \PUCDz 
\right],
\end{equation}
to be determined from available microstructural data. For the sake of
generality, we assume that the microstructure configuration is
characterized by microstructural function associated with~(at most)
warp and weft directions; i.e. functions $\tppF_{\ww}$ and
$\lpF_{\ww}$ for the warp cross-section and descriptors $\tppF_{\wf}$
and $\lpF_{\wf}$ for the weft cross-section,
recall~\figref{SEPUC}(b).\footnote{%
  When the same statistics is assumed for both warp and weft
  directions we set $\PUCDx = \PUCDy$.}
In particular, see also~\cite{Zeman:2007:FRM,Yeong:1998:RRM}, the
following quantities are introduced to measure the similarity between
a SEPUC and the original microstructure:
\begin{eqnarray}
F_{\tppF}^2(\vek{y}) & = & 
\frac{1}{\imax \jmax}\sum_{\phs \in \{ \ww, \wf \}} 
\sum_{i = -\imax}^{\imax}
\sum_{j = -\jmax}^{\jmax}
\left( \tppF_\phs(\vek{y},i,j) 
- 
\orig{\tppF}_\phs( i, j ) \right)^2,
\\
F_{\lpF}^2(\vek{y}) & = & 
\frac{1}{N_d N_\ell}
\sum_{\phs \in \{ \ww, \wf \}} 
\sum_{i = 1}^{N_d}
\sum_{j = 1}^{N_\ell}
\left( 
\lpF_\phs(\vek{y},i,j) 
- 
\orig{\lpF}_\phs( i, j )
\right)^2,
\end{eqnarray}
where, e.g. $\tppF_{\ww}( \vek{y}, i, j )$ denotes the two-point
probability function determined for the warp cross-section of a SEPUC
described by parameters $\vek{y}$ and the value of argument $\vek{X} =
[i,j]$, $\lpF_{\wf}( \vek{y}, i, j )$ stores the value of the
weft-section lineal path function for the $j$-th pixel of the $i$-th
segment, $\orig{\tppF}_\bullet$ and $\orig{\lpF}_\bullet$ denote the
statistics related to original media and the dimensions $\imax$ and
$\jmax$ are determined as half of the minimum height and width of the
bitmaps representing a SEPUC and the reference image.

The two-dimensional data can be complemented by independent
experimental measurements of three-dimensional volume fractions of the
tow phase, see~\cite{Tomkova:2006} for further details. Such
information is accounted for by an additional discrepancy measure
\begin{equation}
F_{\volfrac}(\vek{y}) = | \phi( \vek{y} ) - \orig{\volfrac} |,
\end{equation}
where $\phi( \vek{y} )$ denotes the SEPUC three-dimensional volume
fraction and $\orig{\volfrac}$ is the target value. The former
quantity is determined from the analytical representation of the SEPUC
geometry~\cite{Kuhn:1999:MPWCG} using an adaptive Simpson
quadrature~\cite[Chapter~4]{Press:1992:NRC} with the relative accuracy
set to $10^{-5}$. 

Moreover, the multiple descriptors can be arbitrary combined in the
form of a weighted sum. For example, if all available information is
employed, the objective function attains the form
\begin{equation}\label{eq:F_SL}
F_{\tppF+\lpF+\volfrac}(\vek{y}) 
=
\weight_{\tppF} F_{\tppF}(\vek{y}) 
+ 
\weight_{\lpF} F_{\lpF}(\vek{y})
+
\weight_{\volfrac} F_{\volfrac}( \vek{y} ),
\end{equation}
with $\weight_\bullet$ denoting scale factors used to normalize the
influence of each descriptor, determined from twenty randomly
generated SEPUCs in the current study.

The final term is introduced into the objective function to eliminate
the intersection of the upper-layer and lower-layer tows. To that end,
we compute the overlap $\delta$ as the minimum signed distance between the
upper and lower tow surfaces and introduce the constraint $\delta \geq
0$ via a polynomial exterior penalty:
\begin{equation}\label{eq:objfunc}
f_\descr(\vek{y}) 
= 
\left(
1 +  \frac{\delta_-(\vek{y})}{\PUCh} 
\right)^\beta
F_\descr(\vek{y}),
\end{equation}
where $\delta_-$ denotes the negative part of $\delta$, $\descr$
refers to a particular combination of the descriptors and the value of
exponent is set to $\beta = 3$. Note that this approach was inspired
by work of Collins et al.~\cite{Collins:2009:T3R} related to
high-density polydisperse particulate composites.

Now, the optimal values of the SEPUC parameters can be determined as
the solution to a box-constrained global optimization problem
\begin{equation}\label{eq:opt_problem}
\vek{y} 
\in 
\Argmin_{\vek{l} \leq \vek{z} \leq \vek{u}}
f_\descr(\vek{z}),
\end{equation}
where the lower and upper bounds $\vek{l}$ and $\vek{u}$ are directly
linked to standard deviations in~\tabref{parameters_SEPUC} acquired
from two-dimensional image analysis. A closer inspection reveals that
objective functions~(\ref{eq:objfunc}) are multi-modal and
discontinuous due to the effect of limited bitmap resolution,
cf.~\cite{Zeman:2003:ACM}. Based on our previous experience with
evolutionary optimization, a stochastic optimization algorithm
\code{RASA}~\cite{Matous:2000:AGA,Hrstka:2003:CS}, based on a
combination of a~real-valued genetic algorithm and the Simulated
Annealing method, is used to solve the optimization
problem~\eqref{opt_problem}.

\subsection{SEPUC for multi-layered C/C composite}\label{sec:SEPUC_valid}

This section is concerned with the analysis of the eight-layer C/C composite
represented by the bitmap appearing in~\figref{img_anal_example}(c). First, to
keep the optimization process manageable, the original $2,261\times 861$ bitmap
was down-sampled to a $1,024 \times 390$ image, resulting in the pixel size of
$4.4~\mu$m. In accordance with conclusions of the previous section, all
available information was employed to determine SEPUC parameters, constrained to
the range $\overline{y}_i \pm 3 \sigma_i$, with $\overline{y}_i$ and $\sigma_i$
denoting the mean and the standard deviation of the $i$-th structural parameter
taken from~\tabref{parameters_SEPUC}. Parameters of the algorithm were set to
the identical values as in~\cite[page 65]{Zeman:2003:ACM} and the termination
criterion was set to $50,000$ of $f_{S+L+\phi}$ function evaluations. The
optimization was executed independently twenty times to obtain reliable results.

\begin{figure}[ht]
\begin{minipage}{.575\textwidth}
\centering
\includegraphics{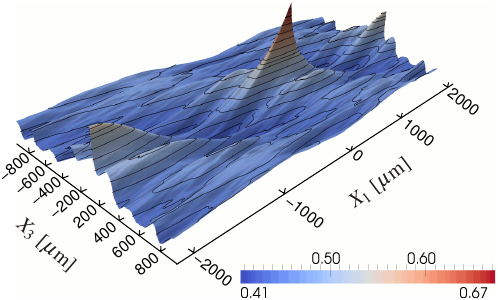}\\
(a)
\end{minipage}
\quad
\begin{minipage}{.4\textwidth}
\centering
\includegraphics{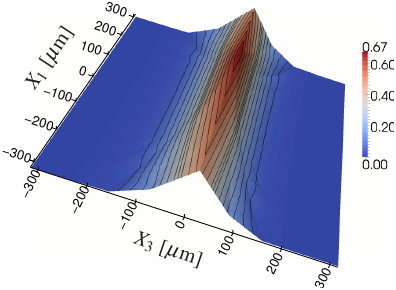}\\
(b)
\end{minipage}

\begin{minipage}{.575\textwidth}
\centering
\includegraphics{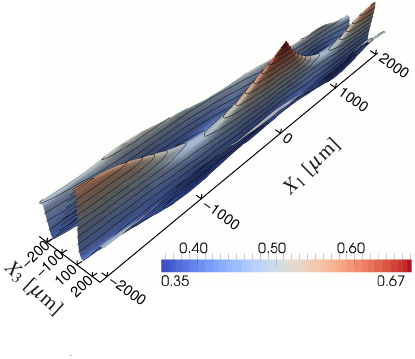}\\
(c)
\end{minipage}
\quad
\begin{minipage}{.4\textwidth}
\centering
\includegraphics{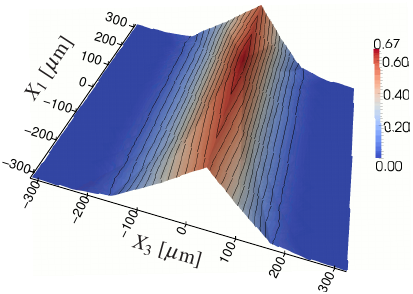}\\
(d)
\end{minipage}
\caption{Statistical descriptors for multi-layered C/C composite and
  SEPUC; (a)~two-point probability function and (b)~lineal path
  function determined for multi-layered composite, (c)~two-point
  probability function and (d)~lineal path function corresponding to
  SEPUC}
\label{fig:orig_SEPUC}
\end{figure}

The resulting two-point and lineal-path functions corresponding to the
SEPUC and the original microstructure appear
in~\figref{orig_SEPUC}. In the $X_1$-axis direction, the original
statistics is well-reproduced, particularly at the $X_3 = 0~\mu$m
plane where the extreme values and the shape of the descriptors are in
almost perfect agreement. The SEPUC also partially captures the local
peaks appearing in the multi-layer system for $X_3 \approx \pm
200~\mu$m, cf.~\figref{orig_SEPUC}~(a). In the perpendicular
direction, we observe that the SEPUC two-point probability function is
influenced by periodic boundary conditions, leading to more
oscillatory behavior and to a slight shift of the minimum values from
$41\%$ to $35\%$. The match between the lineal path functions in terms
of extreme values and shape of the function is even closer, since this
descriptor is non-periodic even for periodic microstructures. Finally
note that the tree-dimensional volume fraction $\phi$ of the SEPUC is
$51\%$, which coincides exactly with the reference value taken
from~\cite{Tomkova:2006}. Therefore, we conjecture that the idealized
SEPUC captures the dominant geometrical features of the original
system; the differences visible from~\figref{orig_SEPUC} arise mainly
due to the periodic boundary conditions and idealized shape of
SEPUC.

In~\tabref{parameters_SEPUC_scaled}, we present the parameters of the SEPUC
together with the standard deviations estimated from independent optimization
runs. The scatter in the identified parameters is mainly induced by
discretization errors and does not exceed three pixel sizes. The values of the
parameters demonstrate that SEPUC captures a moderate horizontal shift of
individual layers and their mutual overlap. These conclusions are further
supported by a three-dimensional representation in~\figref{SEPUC_p2}(b)
and~\figref{izo-rezo-FEM}, showing that SEPUC reproduces the matrix rich regions
together with the strong nesting of individual layers, of course within the
constraints of the selected geometrical model and the tow impenetrability
condition.

\begin{table}[ht]
\caption{Optimal parameters of the two-layer periodic unit cell free of pores; 
  numbers in parentheses correspond to standard deviations}
\centering
\begin{tabular}{rrrrrr}
\hline
$\PUCa$~[$\mu$m] & $\PUCb$~[$\mu$m] & $\PUCg$~[$\mu$m] & 
$\PUCh$~[$\mu$m] & $\PUCDx = \PUCDy$~[$\mu$m] & $\PUCDz$~[$\mu$m] \\
\hline
$2,181$  & $118$ & $394$ & $251$ & $288$ & $-47$ \\
($10.0$) & ($0.3$) & ($0.1$) & ($1.3$) & ($10.5$) & ($2.0$) \\
\hline
\end{tabular}
\label{tab:parameters_SEPUC_scaled}
\end{table}

\subsection{Computational model}~\label{sec:SEPUC_poros}

A crucial step in the successful calculation of effective properties
is the finite element discretization of SEPUC. This step typically
implies the use of conforming finite element
meshes~\cite{Wentorf:1999:AMCW} enabling the implementation of
periodic boundary conditions by assigning the same code numbers to
homologous nodes on opposite sides of a rectangular unit
cell~\cite{Michel:1999:EPC,Zeman:2001:EPG}. Such meshes are also
expected to explicitly resolve all heterogeneities so that only one
phase is present in each finite element. In case of the two layer
SEPUC these include not only the warp and weft tow boundaries but also
the boundary of major voids. Based on the results provided by X-ray
microtomography we assumed disordered voids created by coating each
fiber tow with a layer of matrix such that the volume of coating
complies with the respective volume fraction of the matrix phase. The
remaining space of the total volume of SEPUC then devolves upon the
porous phase.

\begin{figure}[th]
\centering
\begin{tabular}{c@{\hspace{5mm}}c}
\includegraphics[width=.47\textwidth]{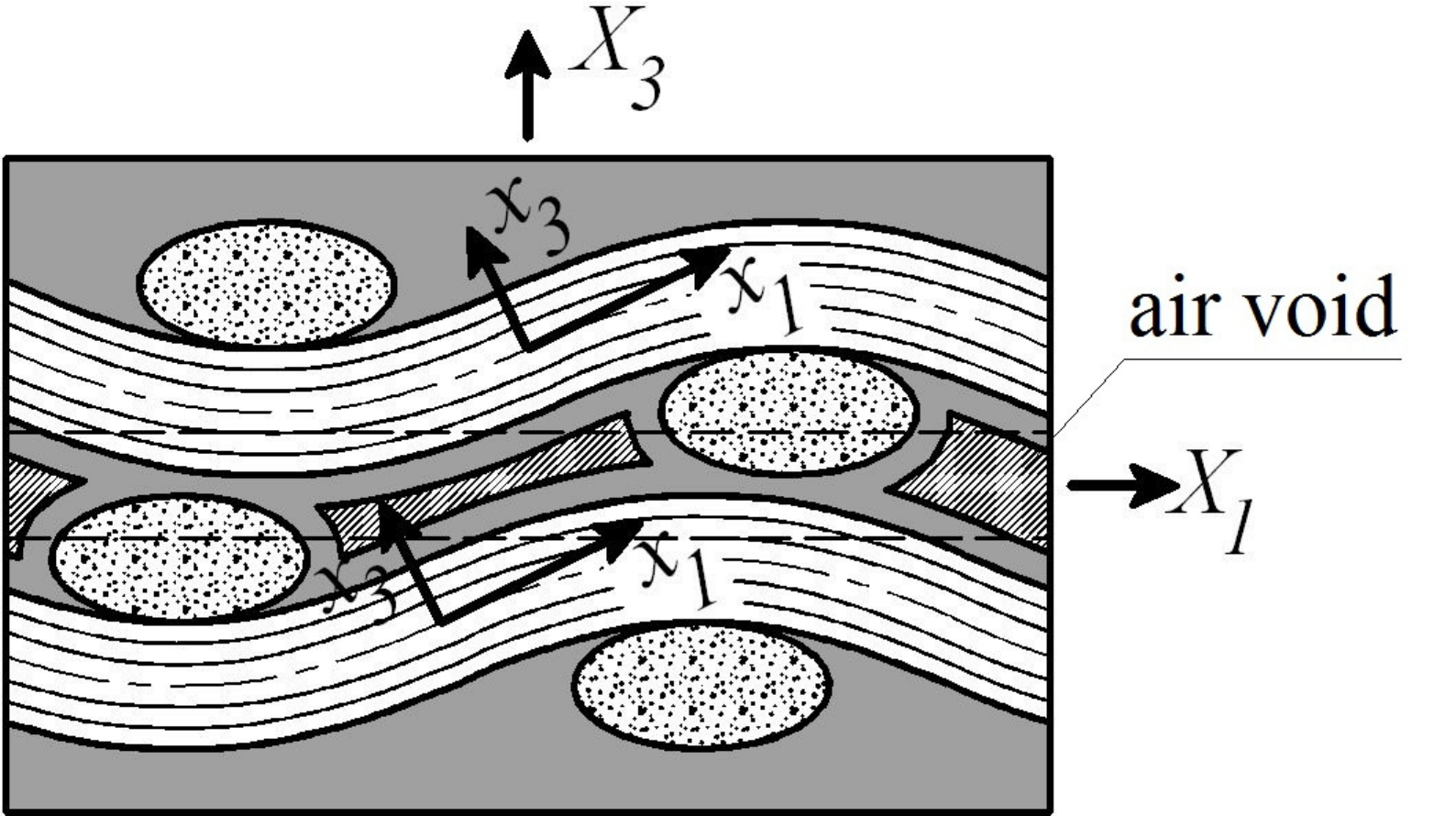} &
\includegraphics[width=.47\textwidth]{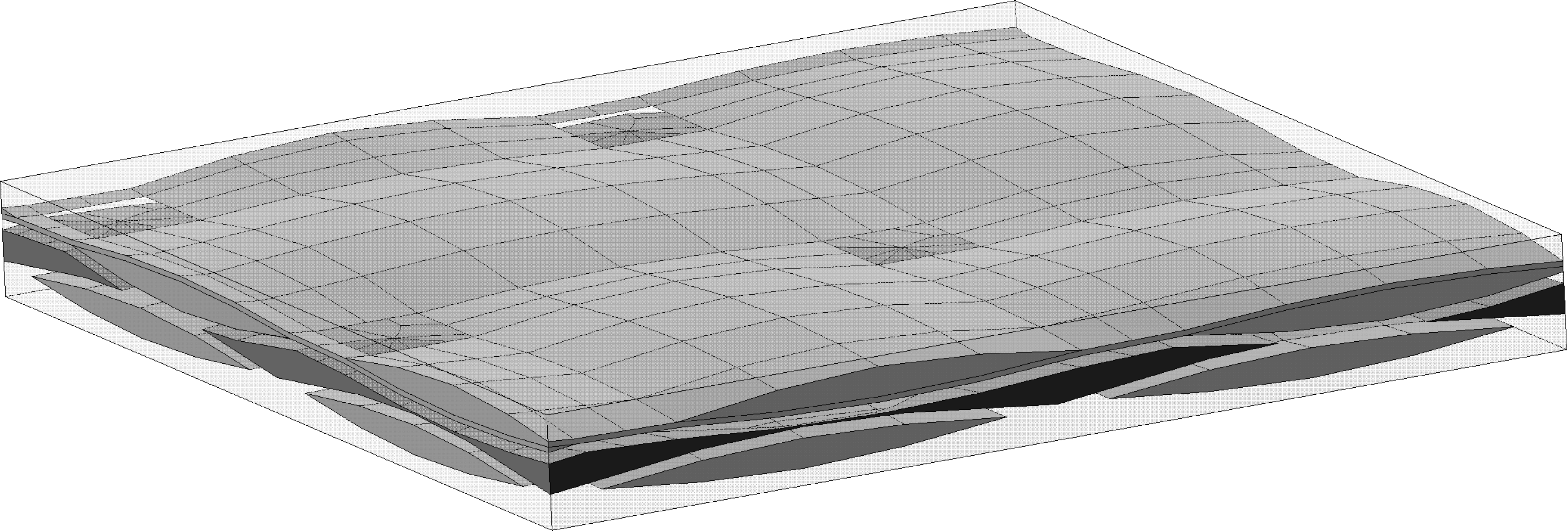} \\
(a) & (b)
\end{tabular}
\caption{
(a)~Two-dimensional cut of a two-layer model with distorted voids
(b)~3D view of the geometry of a two-layer UC model with distorted voids
}
\label{fig:SEPUC_p2}
\end{figure}

As evident from \figref{SEPUC_p2} relatively sharp areas at fiber-tow
crossings may lead to inappropriately distorted elements if attempting
to match all geometrical boundaries. To avoid such complications we
adopt the approach based on the extended finite element method
(X-FEM), which allows us to treat complex geometries relatively
easily. More specifically, the method enables an application of
regular meshes, which do not have to confirm to physical
boundaries. These are captured by enriching the approximation space of
the finite element with embedded interface exploiting the partition of
unity technique. In the present context, the theoretical formulation
described by Mo\"{e}s et al.~\cite{Moes:2003:CAHCMG} will be pursued.

\begin{figure}[ht]
\begin{center}
\includegraphics[width=0.6\textwidth]{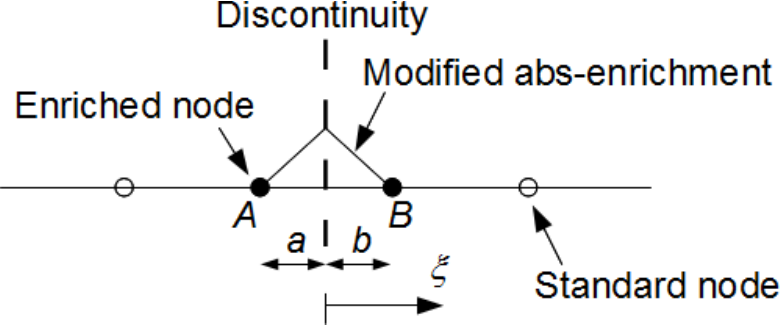}
\caption{Modified abs-enrichment for 1D problem.}
\label{fig:1Denrichment}
\end{center}
\end{figure}

\begin{figure}[th]
\centering
\begin{tabular}{c@{\hspace{2mm}}c@{\hspace{2mm}}c}
\includegraphics[width=.33\textwidth]{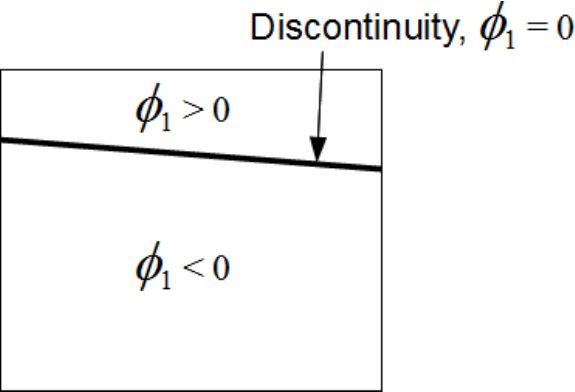} &
\includegraphics[width=.33\textwidth]{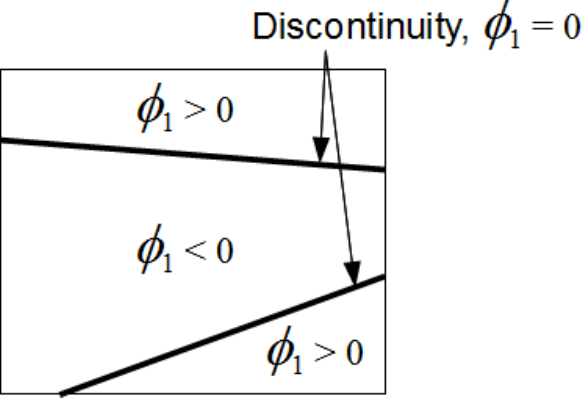} &
\includegraphics[width=.25\textwidth]{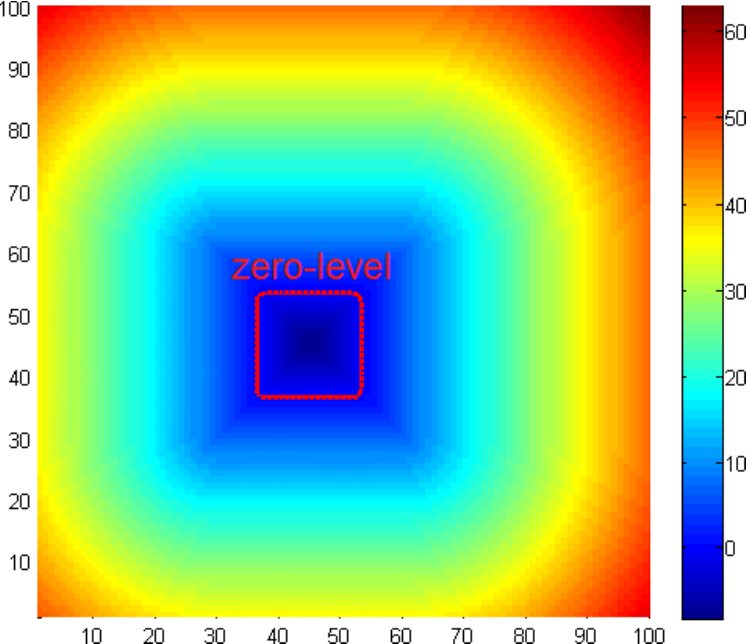} \\
(a) & (b) & (c)
\end{tabular}
\caption{Representation of the level set function function: 
(a) element crossed by a single material interface, 
(b) element crossed by two interfaces of the same material phase,
(c) example of the level set function for a square plate with a hole}
\label{fig:LSFunc}
\end{figure}

To introduce the subject, consider a one-dimensional problem in
\figref{1Denrichment} with an element AB crossed by a material
discontinuity. The principal idea of X-FEM is to augment the standard
approximation space of the displacements or temperatures with a
specific enrichment function $\psi(\M{X})$ which renders the
corresponding gradients discontinuous along the material interface.
Following~\cite{Moes:2003:CAHCMG} and for simplicity limiting
attention to the heat conduction case the augmented approximation of
the temperature field reads
\begin{equation}
{\theta}(\M{X}) = \sum_{i\in I} {N}_i(\M{X}){\theta}_i + \sum_{j\in I\fl}
{N}_j\fl(\M{X})\psi(\M{X}){a}_j,\label{eq:XFEM-1}
\end{equation}
where ${N}_i$ are the standard shape functions, $I$ represents the
total number of finite element nodes in the analyzed domain,
$I\fl\subset I$ gives the number of nodes for which the support is
split by the interface and ${a}_j$ are the additional degrees of
freedom. To properly capture the interface location within an element
Sukumar et al.~\cite{Sukumar:CMAME:2001} applied a level set
representation of surfaces through a level set function
\begin{equation}
\phi(\M{X})=\sum_{i\in J}{N}\fl_i(\M{X})\M{\phi}_i,
\end{equation}
where ${N}\fl_i$ are the shape functions building a local partition of
unity, in most cases ${N}\fl_i={N}_i$. Then, $J$ stands for the number
of nodes of the element for temperature discretization containing the
point $\M{X}$. The nodal values $\M{\phi}_i$ represent the signed
distance of the element node to the interface with either a positive
or a negative value depending on the material to which it belongs as
sketched in \figref{LSFunc}(a). This function then locates interfaces
implicitly as a union of points for which it attains a zero value
(zero-level). An example of a level set function for a square plate
with an embedded hole is shown in \figref{LSFunc}(c). The situation in
which two interfaces of the same material phase are found in a given
element, see \figref{LSFunc}(b), should be avoided since the resulting
approximation would not yield the iso-zero of the level set function
inside the element correctly. This condition thus calls for a minimum
mesh refinement at least locally in the vicinity of two interfacases
of the same material, i.e. interfaces defined by the same level set
function.

Given the level set function, Mo\"{e}s et al.~\cite{Moes:2003:CAHCMG}
introduced a specific form of the enrichment function in the form
\begin{equation}
\psi(\M{X}) = \sum_{i \in J} \left|\phi_i\right| {N}_i\fl(\M{X}) - \left|\sum_{i \in J} \phi_i {N}_i\fl(\M{X})\right|,
\end{equation}
where $\phi_i$ denotes the level set value in the node $i$. Note
that the actual value of $\psi$ along the interface is irrelevant as
long as it captures the weak discontinuity in gradient fields
properly. For the general case of $K$ enrichment we get
\begin{equation}
{\theta}(\M{X}) = \sum_{i=1}^{I} {N}_i(\M{X}) {\theta}_i + \sum_{k=1}^K \sum_{j\in J_k} {N}_j\fl(\M{X}) \psi(\M{X})^k{a}_j^k,\label{eq:XFEM-2}
\end{equation}
where $J_k$ stands for a nodal subset of the enrichment $k$. In the
elasticity case Eq.~\eqref{XFEM-2} applies to each component of the local
displacement field $\M{u}(\M{X})$. Notice that for the analyzed porous
textile composites three enrichment functions might arise for a single
element associated in turn with the weft and warp tows and a void.

Several issues worth of noting arise with numerical implementation. The
first two are general and include proper numerical integration and
application of periodic boundary conditions as discussed
in~\cite{Jerabek:2010:NFMCC} and~\cite{Moes:2003:CAHCMG},
respectively.

\begin{figure}[th]
\centering
\begin{tabular}{c@{\hspace{5mm}}c@{\hspace{5mm}}c@{\hspace{5mm}}c}
\includegraphics[width=.20\textwidth]{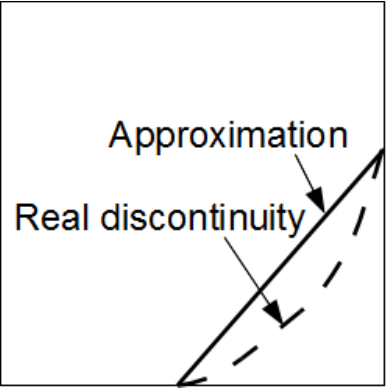} &
\includegraphics[width=.20\textwidth]{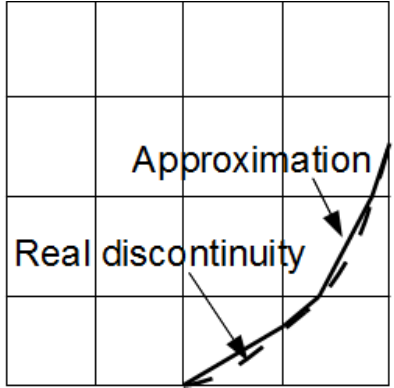} &
\includegraphics[width=.23\textwidth]{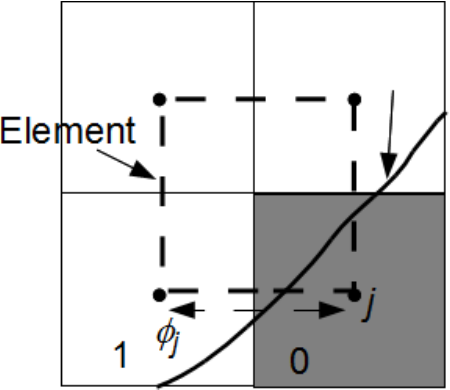} &
\includegraphics[width=.23\textwidth]{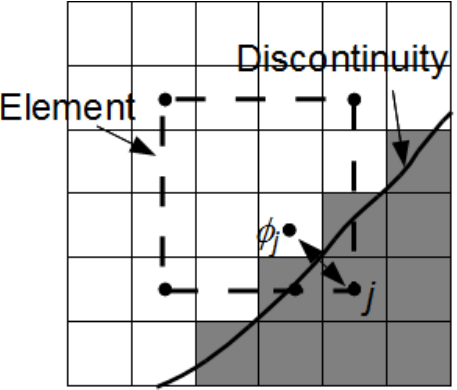} \\
(a) & (b) & (c) & (d)
\end{tabular}
\caption{}
\label{fig:LSapprox}
\end{figure}

Other issues are concerned with the approximation of shape and
location of a real interface. The accuracy depends on the
interpolation of the level set function, typically on the finite
element mesh, and evaluation of the sign distances $\phi_j$. It can be
expected that the finer the mesh the smaller the geometrical error
will be. This is demonstrated in Figs.~\ref{fig:LSapprox}(a)(b) for
the case of a linear interpolation of the level set function which
intrinsically replaces a curved interface by a straight line in 2D or
a plane in 3D problems. To measure the sign distances $\phi_j$, we
adopted a binary representation of the geometrical model instead of
using explicit definition of individual interfaces, which for the
porous phase would be rather complicated.
\begin{figure}
\centering
\begin{tabular}{c@{\hspace{10mm}}c}
\includegraphics[width=.33\textwidth]{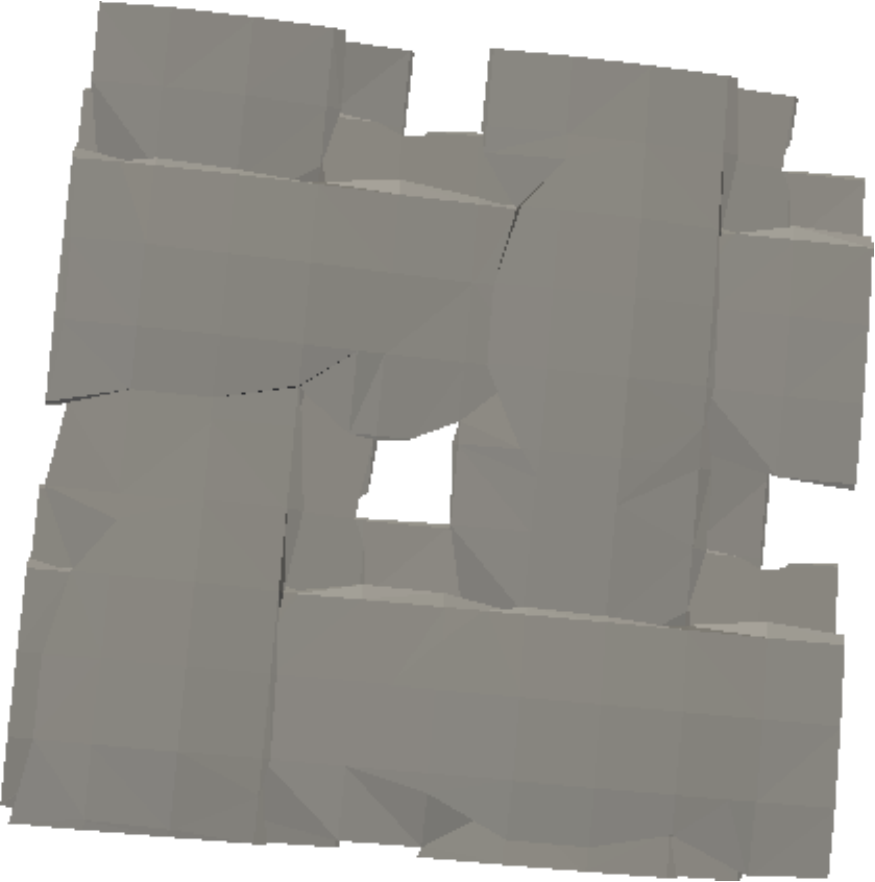} &
\includegraphics[width=.33\textwidth]{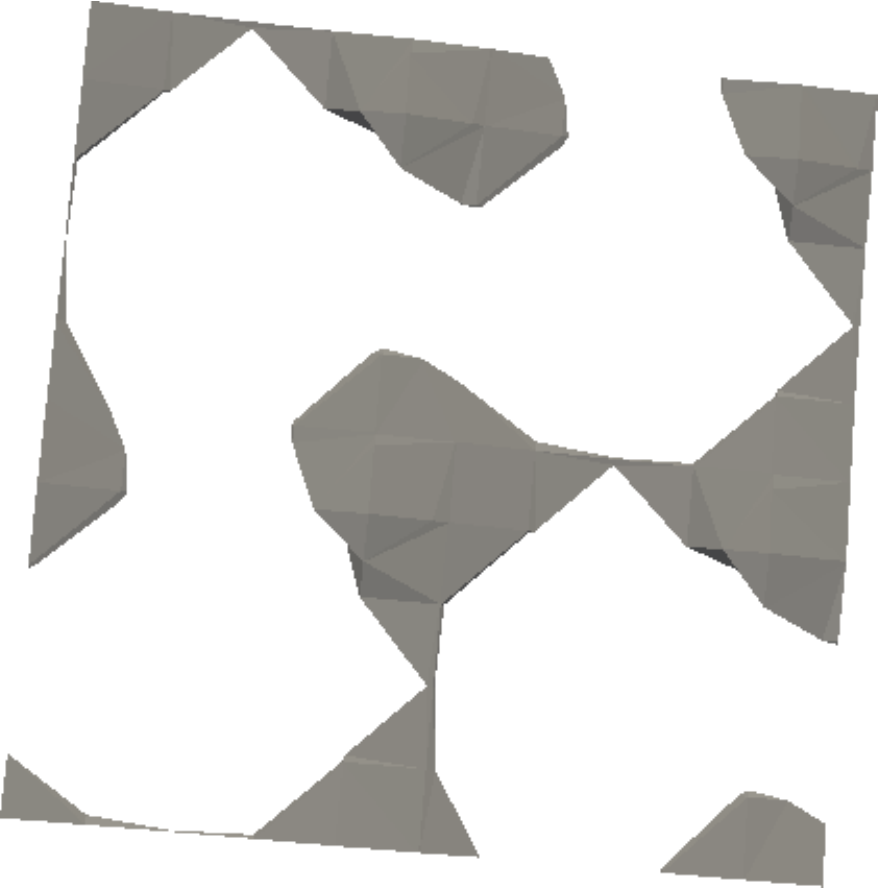} \\
(a) & (b)\\
\includegraphics[width=.33\textwidth]{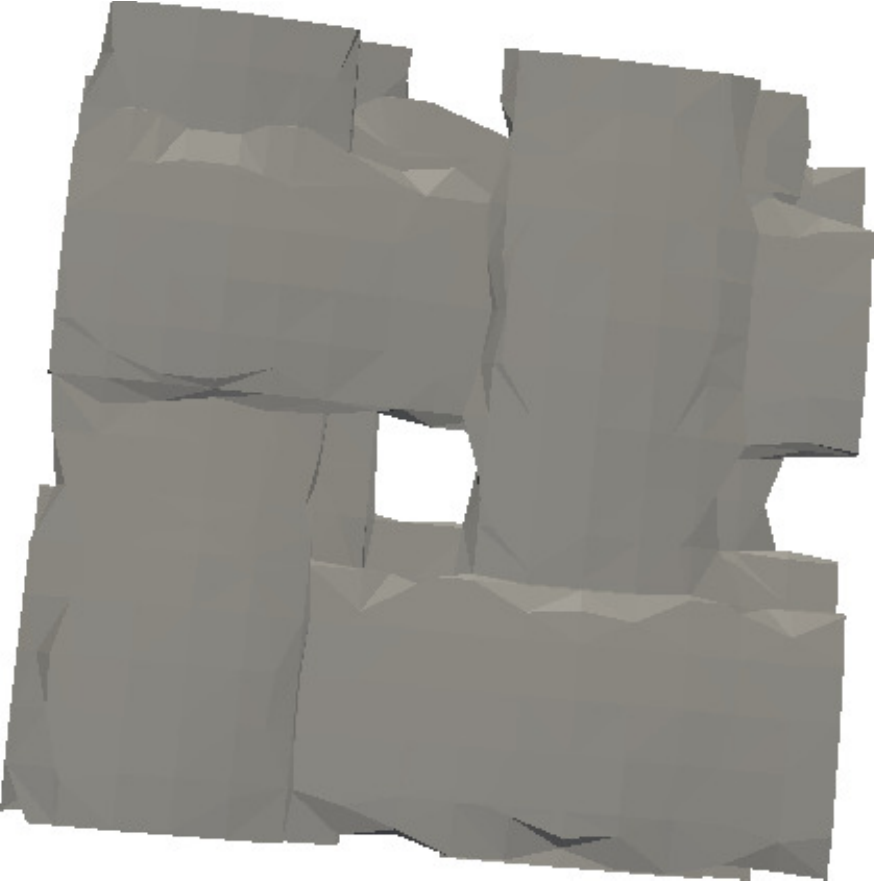} &
\includegraphics[width=.33\textwidth]{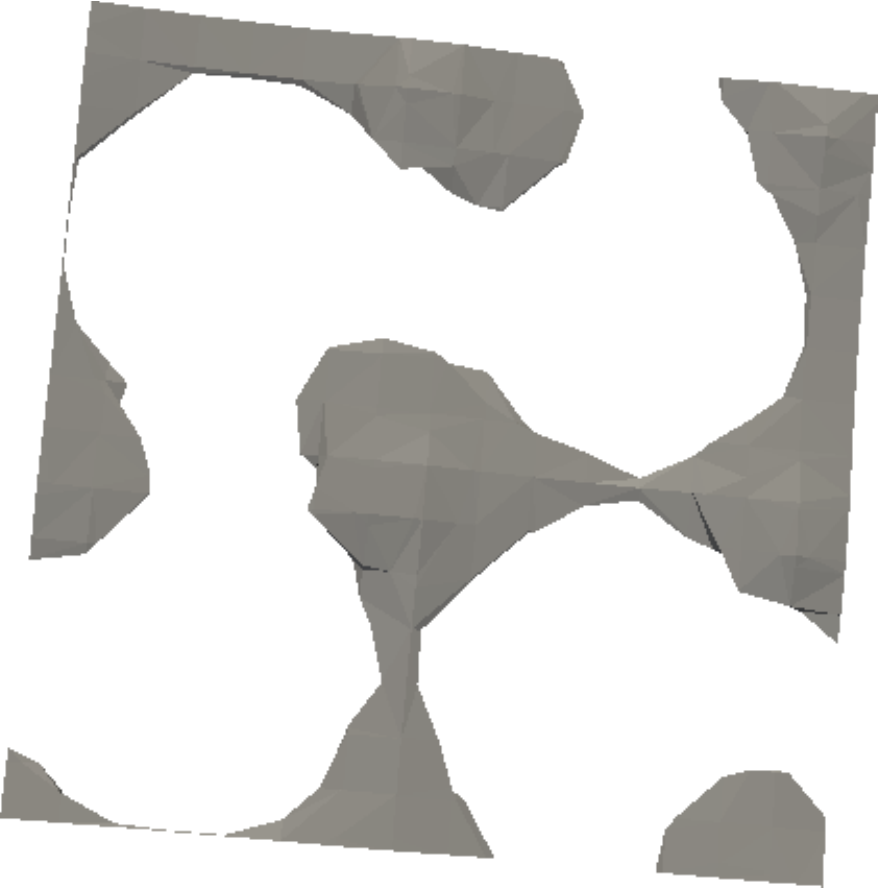} \\
(c) & (d)\\
\includegraphics[width=.33\textwidth]{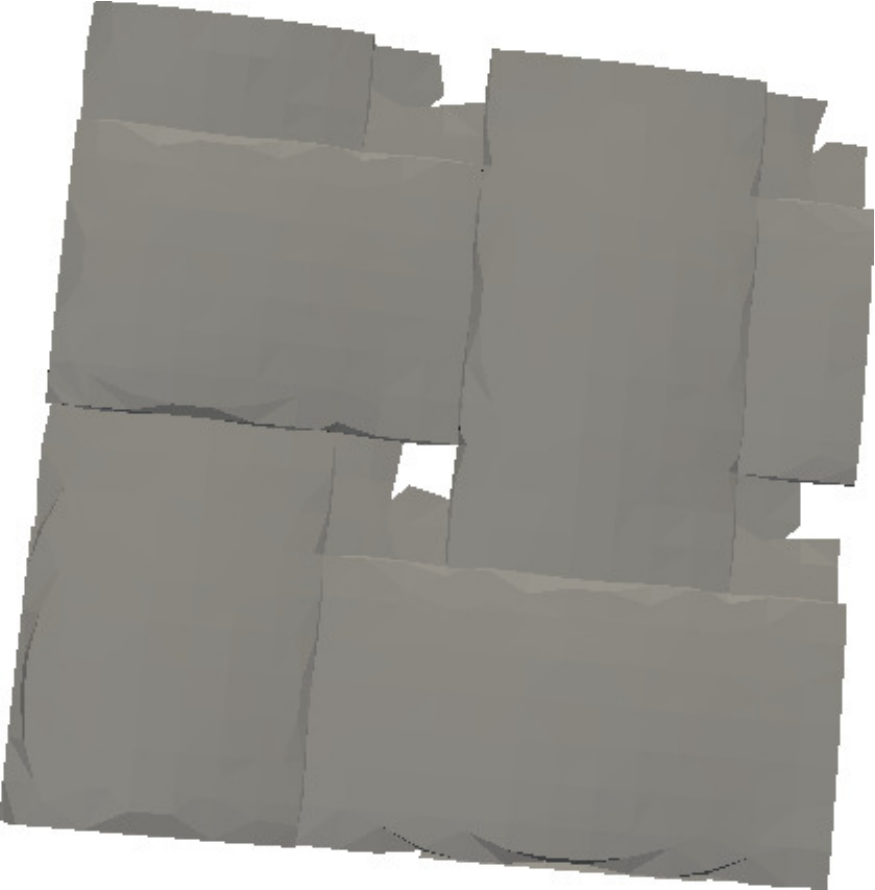} &
\includegraphics[width=.33\textwidth]{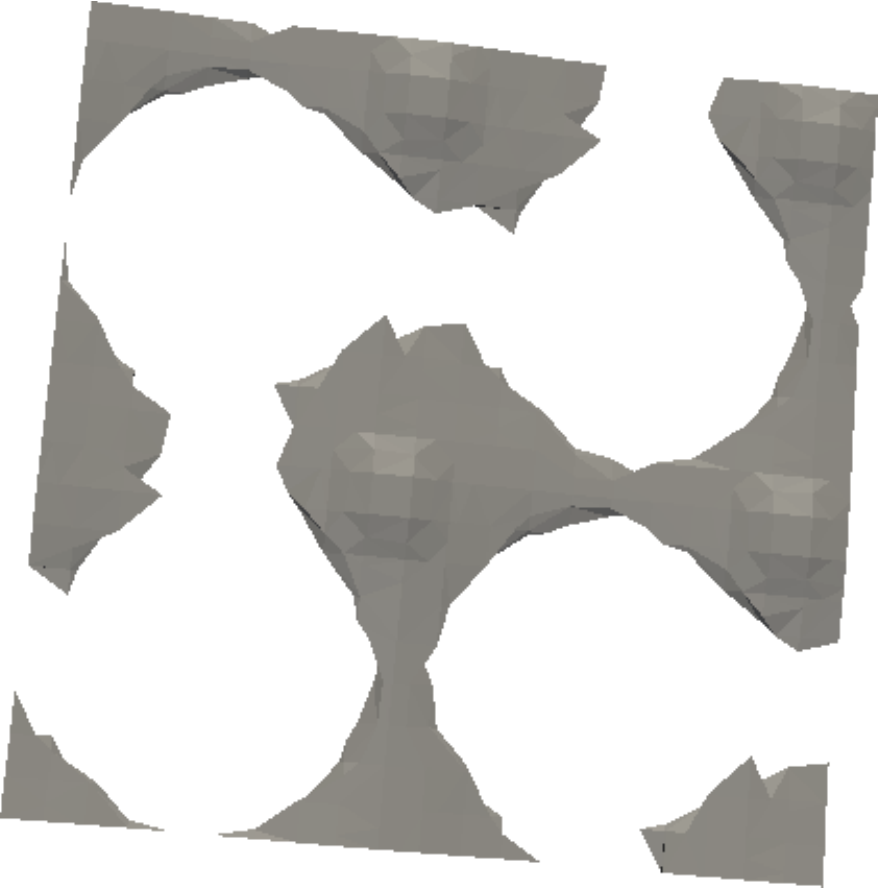} \\
(e) & (f)
\end{tabular}
\caption{Iso-zero representation of fiber-tows and a porous phase of the SEPUC in \figref{SEPUC_p2} based solely on the finite element approximation for three regular meshes: 
(a,b) $10\times{10}\times{8}$, 
(c,d) $15\times{15}\times{10}$,  
(e,f) $20\times{20}\times{15}$}
\label{fig:izo-rezo-FEM}
\end{figure}
The images were created consecutively for the three material phases
(fiber-tow in warp and weft directions and voids) starting with the
resolution corresponding to the underlying uniform finite element
mesh. For each interface the value of $\phi_j$ was defined as a
distance of the zero-value pixel associated with the node $j$ to the
nearest non-zero pixel as seen in \figref{LSapprox}(c). Notice that
element nodes are imagined in the centers of pixels. To receive more
accurate results each cube can be further subdivided into smaller
parts (pixels) as displayed in \figref{LSapprox}(d).

The need for sufficiently fine meshes to accurately locate the
material interfaces becomes particularly important in the light of
volume fractions of individual phases, which play a crucial role in
the actual predictions of effective properties. \figref{izo-rezo-FEM}
displays the iso-zero level set separately for the fiber-tows and the
porous phase for the selected three meshes of different
refinement. One should notice, apart from an improved representation
of the interface location, the change in volume of individual phase.

\begin{table}[ht]
\caption{Mesh properties and associate phase volume fractions}
\centering
\begin{tabular}{l|c|c|c|c}
\hline
Mesh & Conforming & \multicolumn{3}{c}{Uniform (X-FEM)}\\
\cline{3-5} & (FEM) & $20\times{20}\times{15}$ & $15\times{15}\times{10}$ & $10\times{10}\times{8}$\\
\hline
$\#$ elements & 69169 & 6000 & 2250 & 800 \\
$\#$ equations &  11696 & 15148  & 6893  & 2726 \\
$\#$ int.points & 69169 & 182340 &  103076 & 40372\\
\hline
$c_m$ & 0.431& 0.451&	0.495&	0.553\\
$c_f$ & 0.497& 0.484&	0.456&	0.409\\
$c_p$ & 0.071& 0.065&	0.050&	0.038\\
\hline
\end{tabular}
\label{tab:xfem-mesh}
\end{table}

\tabref{xfem-mesh} provides summary on individual finite element
meshes. The corresponding results attributed to a conforming mesh are
also listed for comparison. Clearly, further refinement would be
needed in both the X-FEM and standard formulation to arrive at $8$\%
porosity measured by X-ray microtomography. The actual distribution of
porosity plotted in \figref{xray-CCpore}(b) is, however, captured by
X-FEM approximation relatively well. This issue will yet be commented
on in~\secref{FEM} when discussing the results of numerical analysis.

\section{Numerical evaluation of effective properties}\label{sec:EXAMPLES}

Numerical evaluation of effective elastic moduli and thermal
conductivities, the most classical subject in micromechanics, is
descried in this section in support of the proposed concept of SEPUC
applied to multi-layered C/C composites. This selection of
mechanical and heat conduction problem is promoted not only by
available experimental measurements but also by their formal
similarity, considerably simplifying the theoretical treatment as seen
hereinafter.

\subsection{Theoretical formulation of homogenization}\label{sec:EXAMPLES_theory}

First-order homogenization approaches are now well established and
described in many papers~\cite[to cite a
  few]{Michel:1999:EPC,Kouznetsova:2001:AMMM} to provide estimates of
effective properties of material systems with periodic
microstructures. Bearing in mind the analogy between basic quantities
related to heat conduction problems (i.e. the local microscopic
$\mH(\vek{x})$ and uniform macroscopic $\MH$ temperature gradients and
fluxes $\mQ(\vek{x})$ and $\MQ$ as their conjugate measures) and the
corresponding quantities applied to mechanical problems (local
$\mE(\vek{x})$ and uniform $\ME$ strains and the conjugate stress
measures $\mS(\vek{x})$ and $\MS$), we consider a heterogeneous
periodic unit cell~$\PUC$ and variations of local temperature
$\theta(\vek{X})$ and displacement $\vek{u}(\vek{X})$ fields written
in terms of the uniform macroscopic quantities $\MH$ and $\ME$ as
\begin{equation}\label{eq:local-fields}
\theta(\vek{X}) = \vek{H} \cdot \vek{X} 
+ 
\theta^*(\vek{X}),
\hspace{1.5cm}
\vek{u}(\vek{X}) = 
\vek{E} \cdot \vek{X} 
+ 
\vek{u}^*(\vek{X}),
\end{equation}
where $\theta^{*}$ and $\vek{u}^{*}$ are $\PUC$-periodic temperature
and displacements fluctuations, respectively and $\bullet(\vek{X})$ is
introduced to represent a given quantity in the global coordinate
system $\vek{X}$, recall~\figref{SEPUC}. Next, denoting
$\vek{\chi}(\vek{x})$ the local conductivity matrix and similarly
$\mL(\vek{x})$ the local stiffness matrix, the local microscopic
constitutive equations in the local coordinate system $\vek{x}$ become
\begin{equation}\label{eq:local_const}
\mQ(\vek{x}) = -\vek{\chi}(\vek{x})\mH(\vek{x}),
\hspace{1.5cm}
\mS(\vek{x}) = \mL(\vek{x})\mE(\vek{x}).
\end{equation}
To complete the set of equations needed in the derivation of effective
properties, we recall the Hill lemma for mechanical problem and the
Fourier inequality~\cite{Malvern:1969:IMCM} for an equivalent
representation of steady state heat conduction problem together with
Eqs.~\eqref{local_const} and write the global-local variational
principles, see e.g.~\cite[for further
  details]{Tomkova:2008:IJMCE,Sejnoha:IJES:2007} in the forms
\begin{equation}\label{eq:Hill-Fourier}
\avgs{\delta\mH(\vek{x})\trn\vek{\chi}(\vek{x})\mH(\vek{x})} = 0,\hspace{1.5cm}
\avgs{\delta\mE(\vek{x})\trn\mL(\vek{x})\mE(\vek{x})} = 0,
\end{equation}
where $\avgs{a(\vek{x})}$ represents the volume average of a given
quantity, i.e. $\avgs{a(\vek{x})} = \frac{1}{|\Omega|}\int_\Omega
a(\vek{x}) \de \Omega$. In the framework of finite element
approximation, the discrete forms of local gradients derived from
Eqs.~\eqref{local-fields} read
\begin{equation}\label{eq:local_grad}
\mH(\vek{X}) = \MH+\mtrx{B}^{\theta}(\vek{X})\vek{\theta}^{*}_d,
\hspace{1.5cm}
\mE(\vek{X}) = \ME+\mtrx{B}^{u}(\vek{X})\vek{u}^{*}_d,
\end{equation}
where $\mtrx{B}^{\bullet}$ stores the derivatives of the element shape
functions w.r.t. $\vek{X}$ and $\vek{\theta}^{*}_d$ and
$\vek{u}^{*}_d$ are the vectors of the fluctuation part of nodal
temperatures and displacements, respectively. Substituting
Eqs.~\eqref{local_grad} into Eqs.~\eqref{Hill-Fourier} gives
\begin{eqnarray}\label{eq:HF-dicrete}
\delta\vek{\theta}^{*}_d\trn\avgs{\mtrx{B}^{\theta}(\vek{X})\trn\vek{\chi}(\vek{X})\mtrx{B}^{\theta}(\vek{X})}\vek{\theta}^{*}_d &=& 
-\delta\vek{\theta}^{*}_d\trn\avgs{\mtrx{B}^{\theta}(\vek{X})\trn\vek{\chi}(\vek{X})}\MH,\\
\delta\vek{u}^{*}_d\trn\avgs{\mtrx{B}^{u}(\vek{X})\trn\mL(\vek{X})\mtrx{B}^{u}(\vek{X})}\vek{u}^{*}_d &=& 
-\delta\vek{u}^{*}_d\trn\avgs{\mtrx{B}^{u}(\vek{X})\trn\mL(\vek{X})}\ME,\label{eq:Hill-dicrete}
\end{eqnarray}
to be solved for nodal temperatures $\vek{\theta}^{*}_d$ and nodal
displacements $\vek{u}^{*}_d$. Combining Eqs.~\eqref{local_grad} and
Eqs.~\eqref{local_const} now allows us to write the volume averages of
local heat fluxes and local stresses as
\begin{eqnarray}
\MQ & = & \avgs{\mtrx{T}^{\theta}(\vek{X})\trn\mQ(\vek{x})} \,=\,
-\frac{1}{|\Omega|}\int_\Omega\mtrx{T}^{\theta}(\vek{X})\trn\vek{\chi}(\vek{x})\mtrx{T}^{\theta}(\vek{X})\mH(\vek{X})\de\Omega,\label{eq:Q}\\ 
\MS & = & \avgs{\mtrx{T}^{\mE}(\vek{X})\trn\mS(\vek{x})} \,=\,
\frac{1}{|\Omega|}\int_\Omega\mtrx{T}^{\mE}(\vek{X})\trn\mL(\vek{x})\mtrx{T}^{\mE}(\vek{X})\mE(\vek{X})\de\Omega,\label{eq:S}
\end{eqnarray}
also showing the relationship between material matrices in the local
and global coordinate systems in terms of transformation
matrices $\mtrx{T}^{\theta}, \mtrx{T}^{\mE}$, see
e.g.~\cite{Bittnar:1996:NMM,Zeman:2003:ACM,Tomkova:2008:IJMCE,Vorel:2008:SEM},
\begin{equation}\label{eq:const_local-global}
\vek{\chi}(\vek{X})=\mtrx{T}^{\theta}(\vek{X})\trn\vek{\chi}(\vek{x})\mtrx{T}^{\theta}(\vek{X}), 
\hspace{1.5cm}
\mL(\vek{X})=\mtrx{T}^{\mE}(\vek{X})\trn\mL(\vek{x})\mtrx{T}^{\mE}(\vek{X}).
\end{equation}
The results of Eqs.~\eqref{Q} and~\eqref{S} renders the macroscopic
constitutive laws in the form
\begin{equation}\label{eq:macro_const}
\MQ=-\vek{\chi}^{\sf H}\MH,\hspace{1.5cm}\MS=\ML\ME,
\end{equation}
where $\vek{\chi}^{\sf H}$ and $\ML$ are the searched homogenized
effective thermal conductivity and elastic stiffness matrices,
respectively. In particular, for a three-dimensional SEPUC the
components of the $3\times 3$ conductivity matrix $\vek{\chi}^{\sf H}$
follow directly from the solution of three successive steady state
heat conduction problems. To that end, the periodic unit cell is
loaded, in turn, by each of the three components of $\MH$, while the
other two vanish. The volume flux averages, Eq.~\eqref{Q}, normalized
with respect to $\MH$ then furnish individual columns of
$\vek{\chi}^{\sf H}$. The components of the $6\times 6$ elastic
stiffness matrix $\ML$ are found analogously from the solution of six
independent elasticity problems together with Eq.~\eqref{S}.

\subsection{Numerical simulations}\label{sec:FEM}

In the framework of hierarchical modeling the analysis proceeded in
two steps. First, the effective elastic properties of a porous yarn
were estimated using the Mori-Tanaka~\cite{Benvensite:1987:MTM}
averaging scheme and the local properties stored in~\tabref{param_CC}.
The results appear in~\tabref{lab_ELAST} showing a good agreement with
experimental values taken from~\cite{Cerny:Carbon:2000}. 

\begin{table}[ht]
\caption{Effective elastic properties of unidirectional C/C composite (porous
  fiber tow) and C/C laminate
  moduli~\cite{Cerny:Carbon:2000} given in [GPa]}
\centering
\begin{tabular}{cccc}
\hline
parameter & porous tow (MT) & porous tow (EXP) & C/C laminate (EXP)\\
\hline
$E_{11}$ & 193.8 & $\approx$ 200 & $\approx$ 65\\
$G_{12}$ & 10.3 & $\approx$ 11.5 & $\approx$ 6\\
\hline
\end{tabular}
\label{tab:lab_ELAST}
\end{table}

This encouraging result promoted the application of the Mori-Tanaka
method also in the derivation of effective thermal
conductivities. Both local phase conductivities and the resulting
estimates of effective properties of a porous yarn are available
in~\tabref{lab_HEAT}. Since providing the theoretical details of the
Mori-Tanaka method goes beyond the present scope, we suggest the
interested reader to consult our recent work on this
subject~\cite{Skocek:2007:MT,Vorel:2008:SEM} emphasizing the
applicability of the Mori-Tanaka method also in the field of imperfect
textile composites.

The second step involved application of the 1st order homogenization
discussed in the previous section and the computational model
developed in~\secref{SEPUC_poros}. The corresponding results are given
\tabref{xfem-results}. The table shows how the X-FEM results converge
with the uniform mesh refinement to those attributed to a conforming
mesh.

\begin{table}[ht]
\caption{Phase~\cite{Tomkova:2006}, unidirectional C/C composite
  (porous fiber tow) and laminate effective thermal
  conductivities~\cite{Bohac:2005}}
\centering
\begin{tabular}{lccc}
\hline
material & \multicolumn{3}{c}{Thermal conductivities [Wm$^{-1}$K$^{-1}$]}\\
\cline{2-4} & $\chi_{11}$ & $\chi_{22}$ & $\chi_{33}$ \\
\hline
air & 0.02 & 0.02 & 0.02\\
fiber & 35 & 0.35 & 0.35\\
matrix & 6.3 & 6.3 & 6.3\\
\hline
porous tow (MT) &  24.12 & 1.05 & 1.42\\
\hline
C/C laminate (EXP) & 10 ({$\ww$}) & 10 ({$\wf$}) & 1.6 \\ 
\hline
\end{tabular}
\label{tab:lab_HEAT}
\end{table}

The most pronounced discrepancies, when compared to standard FEM
computations exploiting conforming meshes, clearly follow from an
inaccurate representation of major porosity when coarse meshes are
used. This is also evidenced by plots of the porous phase in
\figref{izo-rezo-FEM} and the corresponding error in the estimated
volume fractions in \tabref{xfem-mesh}. However, given the complexity
of multi-scale analysis, the theoretical predictions when compared to
experimental data listed also in~\tabref{lab_ELAST}
and~\tabref{lab_HEAT} still provide reasonable confidence in the
proposed approach strongly relying on the concept of statistically
equivalent periodic unit cell.

\begin{table}[ht]
\caption{Comparison of X-FEM and FEM predictions for heat conduction and linear elasticity problem}
\centering
\begin{tabular}{c|c|l|c|c|c}
\hline
\multicolumn{2}{c|}{Problem} & Conforming & \multicolumn{3}{c}{Uniform X-FEM}\\
\cline{4-6} \multicolumn{2}{c|}{properties} & \multicolumn{1}{c|}{FEM ({\bf EXP})} & $20\times{20}\times{15}$ & $15\times{15}\times{10}$ & $10\times{10}\times{8}$\\
\hline
Conduction & $\chi_{11}$ & 8.81 ({\bf 10}) & 8.85 & 8.74 & 8.52\\
$[$Wm$^{-1}$K$^{-1}]$ & $\chi_{33}$ & 1.31 ({\bf 1.6}) & 1.81& 2.15& 2.61\\
\hline
Elasticity & $E_{11}$   & 58.75 ({\bf 65}) & 58.53 & 56.73 & 53.53 \\
$[$GPa$]$  & $E_{33}$   & 8.18      & 13.41 & 15.51 & 17.47\\
           & $G_{23}$   & 3.64      & 5.19 & 5.93 & 6.72\\
           & $G_{12}$   & 7.86 ({\bf 6})   & 7.99 & 8.17 & 8.41\\
$[-]$      & $\nu_{23}$ & 0.05      & 0.05 & 0.06 & 0.06 \\
           & $\nu_{12}$ & 0.03      & 0.06 & 0.07 & 0.08\\
\hline
\end{tabular}
\label{tab:xfem-results}
\end{table}

\section{Conclusions}\label{sec:CONCLUSION}

The present article summarizes recent developments in the study of
imperfect carbon-carbon textile composites initiated already in 2004
in part I~\cite{Zeman:2004:HBPWI} of this series. To arrive at the
present stage of understanding of the complex structural response of
these material systems, the machinery of homogenization tools has been
fully exploited.

In view of real material samples impaired by a large amount of flaws
(transverse and delamination cracks, large intertow vacuoles) even
prior to loading, the modeling strategy based on the well known
concept of periodic unit cell may become preferable, particularly if
the response of the material exceeding its elastic
limit~\cite{Fish:1999:CDM,Fish:00:IJCSE,Herb:CPA:10} is the primary
interest. Presently, its potential is seen mainly in the formulation
of statistically equivalent periodic unit cell that attempts to
accommodate the most severe imperfections of the real
material. Supported by the results derived in the course of this work,
the two-layer SEPUC enhanced by incorporating the porous phase appears
to be a suitable candidate for the computational model of plain weave
textile reinforcement based composites including, apart from the
investigated C/C composites, a large group of textile reinforced
ceramics with their anticipated application in bio-medicine.

\section*{Acknowledgments}
The financial support provided by the GA\v{C}R grants No.~106/08/1379 and
No.~105/11/0224 is gratefully acknowledged. The work of Jan Zeman was also
partially supported by the European Regional Development Fund in the
IT4Innovations Centre of Excellence project (CZ.1.05/1.1.00/02.0070). We extend
our personal thanks to Doc. Ond\v{r}ej Jirou\v{s}ek from the Institute of
Theoretical and Applied Mechanics, Academy of Sciences of the Czech Republic,
for providing the X-ray images and to Doc. Ji\v{r}\'{\i} N\v{e}me\v{c}ek from
the Department of Mechanics of the Czech Technical University in Prague, for
providing the results from nanoindentation tests.

\end{document}